\documentclass{vldb}
\usepackage{times, algorithm, algorithmic}

\usepackage{url}
\usepackage{subfigure}
\usepackage{color}
\usepackage{amsmath}
\usepackage{ntheorem}
\usepackage{cite}
\usepackage{xspace}
\usepackage{listings}
\usepackage{upquote}
\usepackage{wrapfig}
\algsetup{linenosize=\small}

\newenvironment{myitem}{
\begin{itemize}
\setlength{\parskip}{0pt}
\setlength{\itemsep}{0pt}
\setlength{\partopsep}{0pt}
\setlength{\parskip}{0pt}
\setlength{\topsep}{0pt}
\setlength{\parsep}{0pt}}{\end{itemize}
}
\usepackage[table]{xcolor}

\lstdefinelanguage{graphRelationalLang} {
  morekeywords={CREATE, GRAPH, VIEW, VERTEXES, EDGES, ID, FROM, SELECT, WHERE, TO, DIRECTED, UNDIRECTED, FUNCTION, RETURNS, WITH, JAR, CLASS, String, TREE, HASH, INDEX, ON, ASC, ORDER, BY, TOP, AND, OR, IN, HINT, NOT, TEMPGRAPH}
}

\newcommand{\ourSys}{GRFusion\xspace}

\newcommand{\rApproach}{\textit{Native Relational-Core}\xspace}
\newcommand{\gApproach}{\textit{Native Graph-Core}\xspace}
\newcommand{\ourApproach}{\textit{Native G+R Core}\xspace}

\begin{document}
\title{
Empowering In-Memory Relational Database Engines with Native Graph Processing}

\numberofauthors{1}
\author
{\alignauthor
Mohamed S. Hassan, Tatiana Kuznetsova, Hyun Chai Jeong\\ Walid G. Aref, Mohammad Sadoghi\\
\affaddr{Purdue University, West Lafayette, IN}\\
\email{{\large\{}msaberab,~tkuznets,~jeong3,~aref,~msadoghi{\large\}}@cs.purdue.edu}
}

\maketitle

\abstract
\sloppypar
The plethora of graphs and relational data give rise to many interesting graph-relational queries in various domains, e.g., finding related proteins satisfying relational predicates in a biological network. The maturity of RDBMSs motivated academia and industry to invest efforts in leveraging RDBMSs for graph processing, where efficiency is proven for vital graph queries. However, none of these efforts process graphs natively inside the RDBMS, which is particularly challenging due to the impedance mismatch between the relational and the graph models.
In this paper, we propose to treat graphs as first-class citizens inside the relational engine so that operations on graphs are executed natively inside the RDBMS. We realize our approach inside \textit{VoltDB}, an open-source in-memory relational database, and name this realization \ourSys{}. The SQL and the query engine of \ourSys{} are empowered to declaratively define graphs and execute cross-data-model query plans formed by graph and relational operators, resulting in up to four orders-of-magnitude in query-time speedup w.r.t. state-of-the-art approaches.

\section{Introduction}
\label{sec:Introduction}

\begin{figure*}[ht]
\centering
  \subfigure[\rApproach{}]{\includegraphics[width=2.1in]{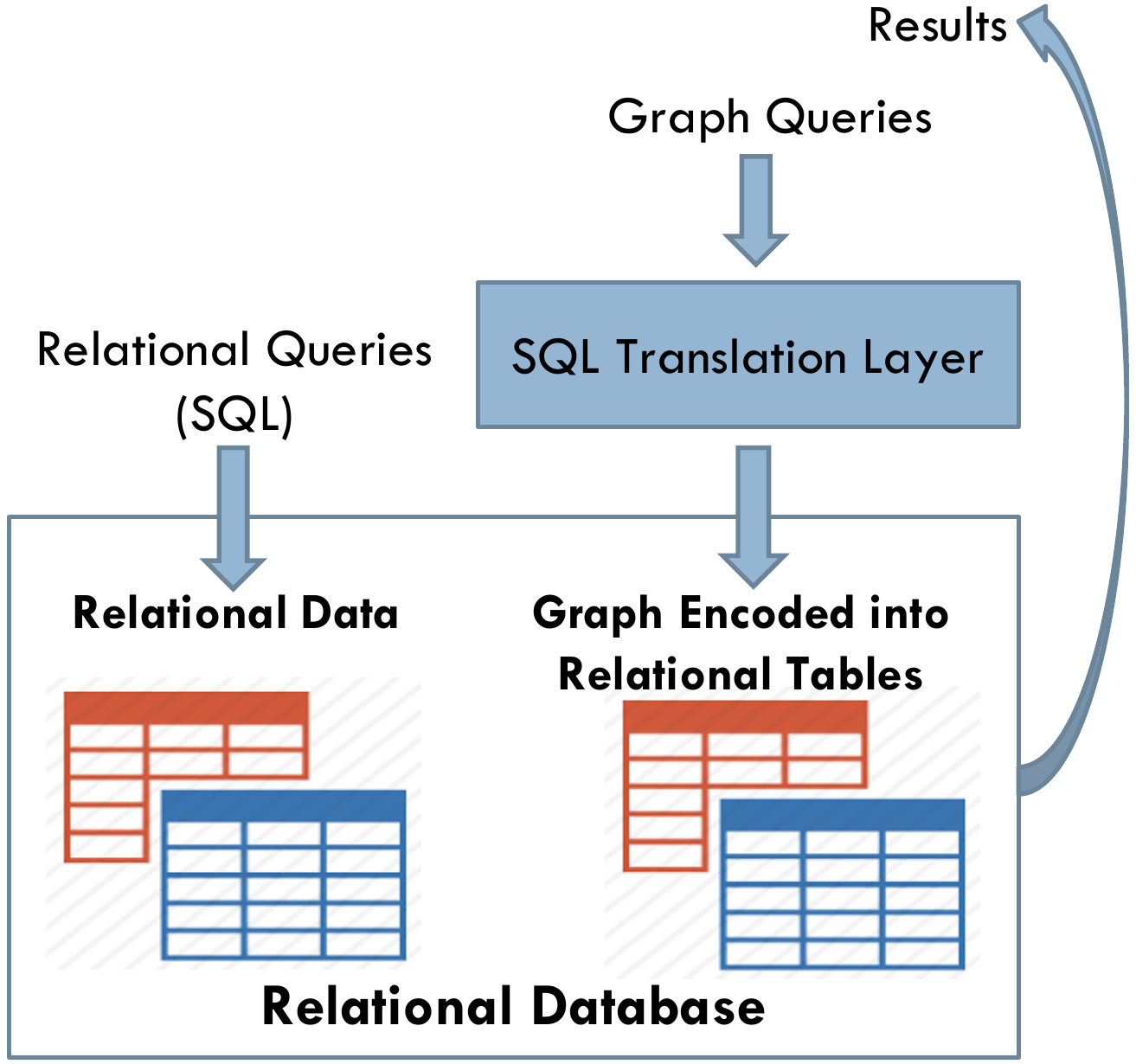}
	\label{Fig:Sys_Embed}
	}
  \hspace{0.1in}
\subfigure[\gApproach{}]
{\includegraphics[width=2.1in]{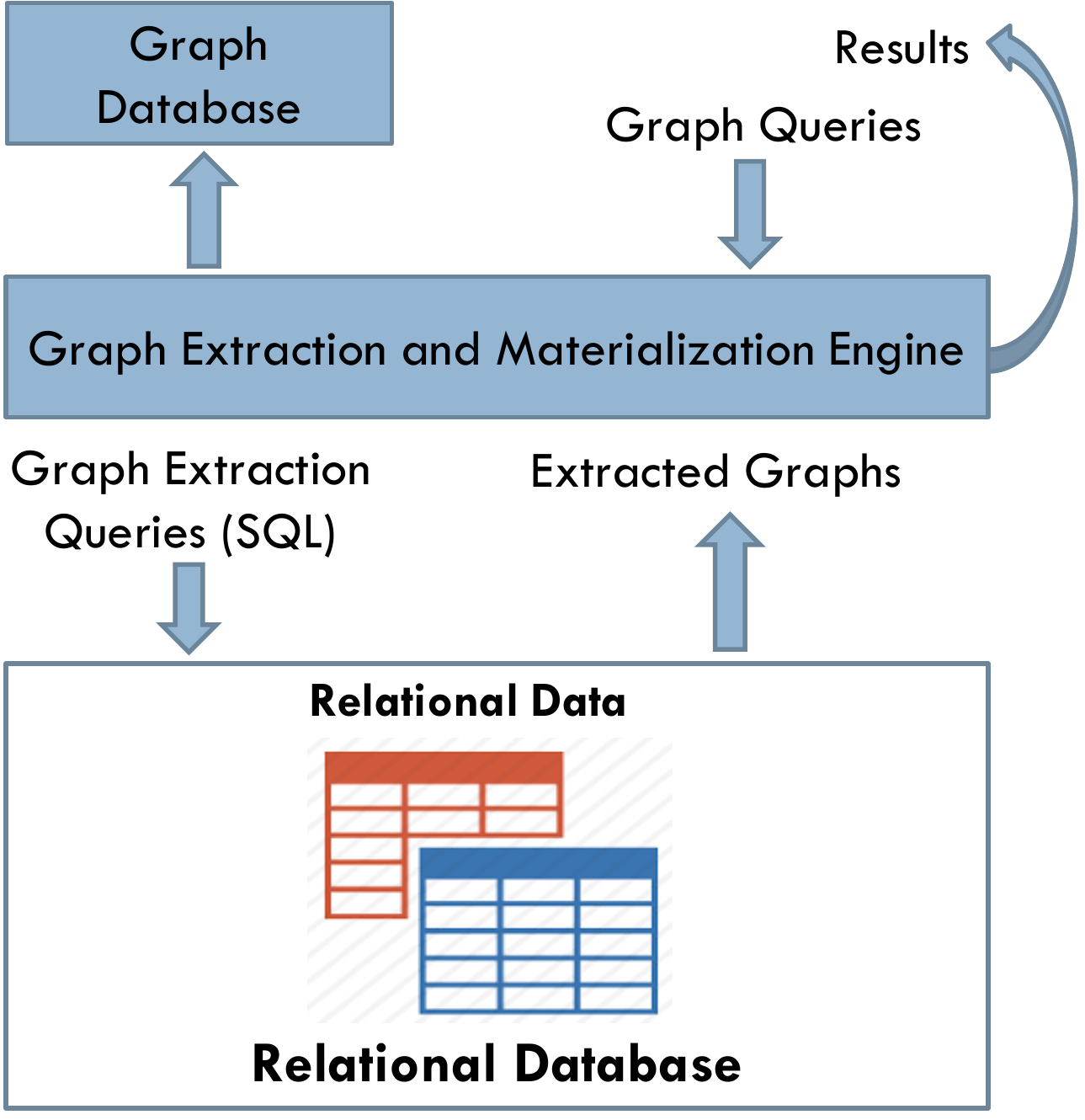}	\label{Fig:Sys_Extract}}
\hspace{0.1in}
\subfigure[\ourApproach{}]
{\includegraphics[width=2.1in]{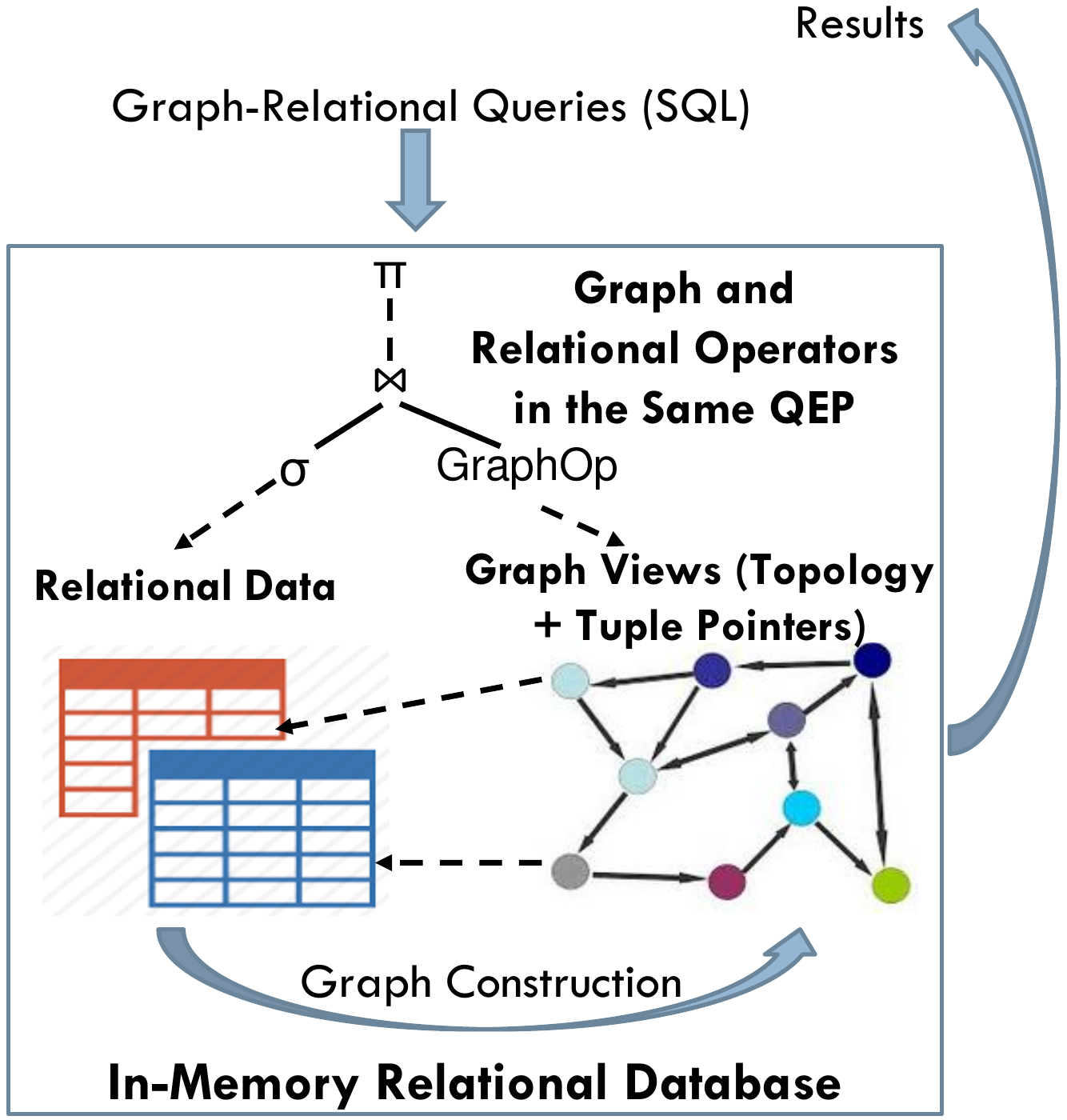}	\label{Fig:Sys_HybridQEP}}
\caption{ 
The various approaches for leveraging relational databases in support of graph processing.}

\label{Fig:Edge_Queries}
\end{figure*}

Graphs are ubiquitous in various application domains, e.g., social networks, road networks, biological networks, and communication networks. The data of these applications can be viewed as graphs, where the vertexes and 
the
edges have relational attributes~\cite{SQLGraph_SIGMOD_2015}, or as traditional relational data with latent graph structures~\cite{GraphGen_Paper_SIGMOD_17}.
Applications would issue queries that reference 
the topology of the graphs along with the data associated with the vertexes and the edges.
For instance, 
a user may be interested to find the shortest path over a road network while restricting the search to certain types of roads, e.g., avoiding toll roads.
In an RDBMS, the filtering predicates can be expressed as relational predicates. 
We refer to these queries as graph-relational queries (or G+R queries, for short). 
G+R queries
have two main ingredients: 1)~graph operations, e.g., shortest-path computation, and 2)~relational predicates or relational sub-queries, e.g., filtering out specific vertexes (or edges).
Moreover, many graphs have relational schemas that describe the data associated with their vertexes and edges. These graphs are the main focus of this work. For instance, the popular STRING biological dataset~\cite{StringDB} has a publicly available relational schema.\footnote{http://string-db.org/download/database.schema.v10.pdf} Another example is the public OpenStreetMap road network~\cite{OpenStreetMap}, where the relational schema is publicly available for different relational databases. Associating
a graph with a relational schema gives rise to many interesting graph-relational queries. For instance, the results of the graph operation may be joined with other relational data to provide insightful results.

Many specialized graph engines with powerful query languages have been 
introduced, 
e.g., Neo4j~\cite{Neo4j} and Titan~\cite{Titan}. However, these systems are not as pervasive and mature as the widely-spread relational databases. Accordingly, various approaches for using an RDBMS to manage graph data have been 
proposed, 
e.g., Grail~\cite{Grail_CIDR_2015} and Aster~\cite{Aster6_VLDB_2014}.
Using 
an RDBMS 
to manage graphs is 
a promising approach
for two main reasons. First, various real scenarios depend on an RDBMS, where the relational data have latent graph structures that may be useful to analyze~\cite{GraphGen_Paper_SIGMOD_17}. Second, graph-relational queries are ubiquitous, and an RDBMS is very powerful in processing the relational constructs of these queries~\cite{Vertica_BigData,SQLGraph_SIGMOD_2015,Aster6_VLDB_2014}. The main challenges when leveraging relational databases in processing graph queries arise from the impedance mismatch between the 
relational and the graph data models.

There are two main approaches for leveraging relational databases in graph query processing, where 
both approaches
share the idea of building an application on top of an RDBMS without modifying the internals of the RDBMS. We refer to these approaches as \rApproach{} and \gApproach{}.
In this paper, we propose and investigate a hybrid approach that we term
\ourApproach{} that exploits the strengths of the former two approaches.
The \rApproach{} approach (e.g., as in SQLGraph~\cite{SQLGraph_SIGMOD_2015} and Grail~\cite{Grail_CIDR_2015}) embeds a graph 
inside of 
relational tables of specific schema.
Then, an application on top of the RDBMS is built to translate specific types of graph queries into SQL statements for the RDBMS to execute. For example, Grail
can translate 
shortest-path queries to procedural SQL~\cite{Grail_CIDR_2015}, while SQLGraph translates Gremlin queries with some restrictions~\cite{Gremlin} into SQL queries~\cite{SQLGraph_SIGMOD_2015}.
Figure~\ref{Fig:Sys_Embed} illustrates the general architecture of the 
\rApproach{} approach.
Notice that the \rApproach{} approach is limited by design to specific types of graph queries.
Although many graph queries and algorithms are hard to translate 
into
SQL statements, tools can be developed 
to automate the translation. 
However, the main issue of the \rApproach{} approach is that the graph operations are evaluated by 
a sequence of 
relational operations (e.g., self-joins) that may be more expensive than traversing a native graph representation.
Moreover, the \rApproach{} approach does not guarantee an easy-to-comprehend relational schema of the embedded graphs in an RDBMS, e.g., the storage-optimized relational schema generated automatically by SQLGraph is hard for users to understand and 
write ad-hoc graph-relational queries~\cite{SQLGraph_SIGMOD_2015}.

The second approach, namely \gApproach{} (e.g., as in Ringo~\cite{Ringo_SIGMOD_15}, GraphGen~\cite{GraphGen_Demo_VLDB_15,GraphGen_Paper_SIGMOD_17}), assumes that graphs are already stored in an RDBMS, where an application on top of the RDBMS is built to extract these graphs to analyze them outside the realm of the RDBMS. This approach follows the same philosophy 
as that of specialized graph databases, where an RDBMS has nothing to do with query execution.
For instance, GraphGen~\cite{GraphGen_Paper_SIGMOD_17} is an application on top of an RDBMS with two main functionalities: 1)~Allow end-users to write Datalog-Like statements to extract graphs from a relational databases, and 2)~Materialize the extracted graphs efficiently in 
main-memory for the end-users to analyze through Graph APIs.
Figure~\ref{Fig:Sys_Extract} illustrates the general architecture of the \gApproach{} approach, where the extracted graphs can also be imported to graph databases for advanced graph querying~\cite{GraphGen_Paper_SIGMOD_17}.
Notice that the RDBMS is not involved in this approach except at the stage of extracting the graphs, where the extraction commands are translated into pure SQL statements by the application on top of the RDBMS. 
Hence, users cannot issue declarative graph-relational queries that reference both the extracted graphs and any other relational data in the RDBMS. This limits the type of queries a user can issue on the owned data. Moreover, if the relational tables storing a graph in the RDBMS are updated, the graph should be re-extracted.
One solution to allow graph-relational queries to reference both the extracted graphs and other relational data is to build another layer that queries both sources. This solution is similar to 
that of Teradata Aster~\cite{Aster6_VLDB_2014}, where a data movement fabric and two different query executors (i.e., a relational executor and a graph executor) are used in processing graph-relational queries. 
However,
integrating the results impose additional overhead
as the results from the graph and the relational executors need to be 
integrated to form the final output. In summary, the \rApproach{} and the \gApproach{} 
approaches
use a vanilla RDBMS. Thus, graphs are recognized as pure relational data by the RDBMS, where no optimizations can consider the graph nature of that relational data.
However, if the necessary layers of the RDBMS are modified to treat graphs as first-class citizens, processing and managing graphs will be more efficient. 

In this paper, we investigate a third approach, namely \ourApproach{}, where graphs are recognized as first-class citizens inside an RDBMS. 
We address the impedance mismatch between the graph and the relational model, where we realize the
\ourApproach{} approach in a centralized version of VoltDB~\cite{VoltDBGitHub,VoltDBCommercial}, the open-source implementation of the H-Store in-memory relational DBMS~\cite{HStore}. In-memory data management is a promising trend with continuous research contributions~\cite{Hekaton, Ringo_SIGMOD_15,GraphGen_Paper_SIGMOD_17,Mem01_MEMSCALE,Mem02_Pipelining,Mem03_ScalingMulticore,Mem04_Logging,Mem05_BTree,Mem06_PlacementInCloud,Mem07_SLACID}. We refer to our realization of this approach as \ourSys{}. The main idea of \ourSys{} is to natively process graphs inside an RDBMS by combining the \rApproach and the \gApproach approaches under the same umbrella. \ourSys{} realizes this idea by separating the graph topology from the relational data associated with the vertexes and the edges, and by proposing graph operators to process the graph topology inside the
RDBMS, where the graph operators 
seamlessly co-exist with other relational operators in the same query execution pipeline (or QEP, for short). A graph topology in \ourSys{} is realized as a native graph structure, where each vertex or edge has pointers to the relational tuples describing their attributes. Hence, a graph topology in \ourSys{} can be viewed as a traversal index of the relational tuples of the vertexes and the edges. In short, \ourSys{} presents cross-data-model QEPs, where the inputs to the QEPs can be either relational data or native graph structures. Figure~\ref{Fig:Sys_HybridQEP} 
illustrates the general idea of the \ourApproach{} approach. First, the end-user provides a declarative statement to create graph views that are initialized from relational data, where a graph view is 
materialized as a new database object. Second, the user is allowed to query the graph views as well as other relational tables or views in the same query.
This paper presents the \ourApproach{} approach, and its realization, \ourSys{}, where we focus on graph traversal queries.
The objective of this paper is not to compete and replace the specialized graph systems. However, the main objective is to empower the pervasive relational databases to support graph traversal queries natively and efficiently. Consequently, the relational-data owners can process important class of graph queries through their RDBMS systems without the cost and the overhead of migrating their data and manage it in a separate different graph system.
The contributions of this paper are as follows:

\begin{myitem}
\item Introducing graphs as native database objects inside an open-source relational database system, namely VoltDB
(Section~\ref{sec:GraphViews}), where online graph updates
are
supported (Section~\ref{sec:GraphUpdates}).
\item Allowing users to seamlessly query and operate on graphs and relational tables simultaneously and declaratively without leaving the realm of the relational database system
(Section~\ref{sec:PathsConstruct}).
\item Introducing graph operators for graph traversals (Section~\ref{sec:GraphOperators}), and showing their ability to seamlessly co-exist with the relational operators to construct cross-data-model query execution pipelines (Section~\ref{sec:HybridQEP}).
\item Addressing the impedance mismatch between the graph model and the relational model w.r.t. query evaluation (Section~\ref{sec:ConceptualEvaluation}).
\item Conducting an extensive performance study of \ourSys{} w.r.t. state-of-the-art systems, and reasoning about the performance benefits of processing graphs in a graph-native representation inside an RDBMS. We compare to SQLGraph, Grail, Neo4j, and Titan, where \ourSys{} achieves up to four orders-of-magnitude query-time speedup (Section~\ref{sec:ExperimentalEvaluation}).
\end{myitem}

The rest of this paper proceeds as follows. Section~\ref{sec:Overview} presents an overview of \ourSys{}. Section~\ref{sec:GraphViews} presents how graph views are defined in \ourSys{}, and how they support online updates for dynamic graphs.
Section~\ref{sec:PathsConstruct} introduces the PATHS construct in \ourSys{} and how it allows declarative graph-traversal queries as well as pattern matching queries. Section~\ref{sec:QueryProcessing} discusses how \ourSys{} evaluates graph-relational queries. Section~\ref{sec:QueryOptimization}
introduces 
some query 
optimization
techniques in \ourSys{}. Section~\ref{sec:ExperimentalEvaluation} presents the experimental evaluation of \ourSys{}.
The related work is discussed in Section~\ref{sec:RelatedWork}, and Section~\ref{sec:Conclusion} contains concluding remarks.

\section{Overview of \ourSys{}}
\label{sec:Overview}

\begin{figure}[h]
\centering
\includegraphics[height=2.3in]
{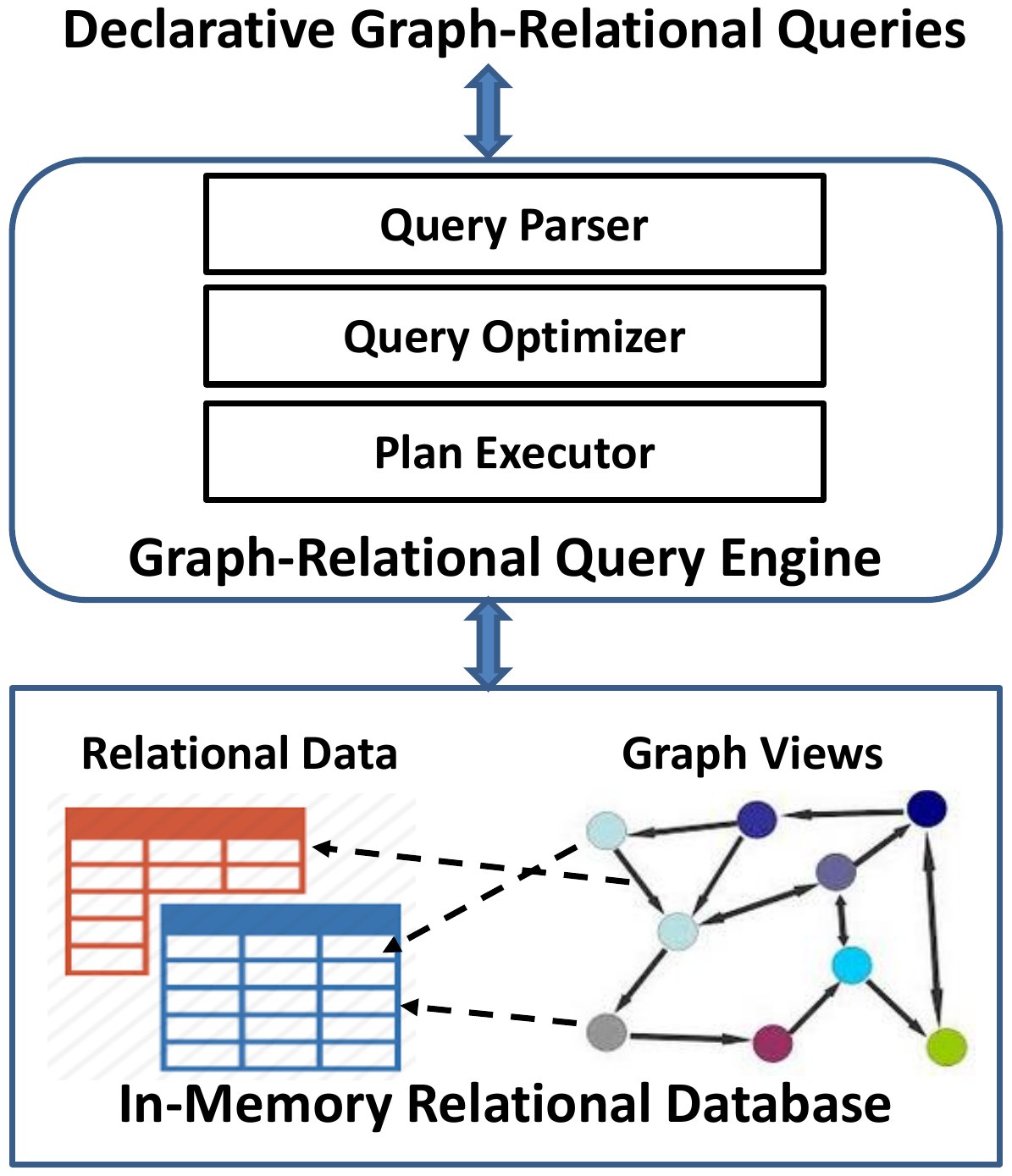}
\caption{\ourSys{}'s architecture allows the query engine to view and process data in both the relational and the graph models.}
\label{Fig:Overview}
\end{figure}

Refer to Figure~\ref{Fig:Overview}. In \ourSys{}, graphs are assumed to be initially stored in relations. In the simplest case, a relational table may have a row for each vertex, and another table could have a row for each edge. Also, the vertexes or the edges data can be obtained through a relational materialized view that joins or filters multiple relational tables. To allow flexibility, \ourSys{} provides the user with a declarative language to define and query graphs. A graph is defined in \ourSys{} by what we term $graph~views$. A $graph~view$ identifies the relational tables or the relational views 
that store the attributes of the vertexes and edges, namely, the \textit{vertexes relational-source}, and the \textit{edges relational-source}, respectively. 
$Graph~views$ define a view of the relational data in the graph model and materializes the graph topology in main-memory in native graph data structures.
The materialized graph topology has a \textit{native graph representation} 
that holds
pointers (e.g., tuple identifiers) to the relational data that describe the vertexes and the edges.
The main idea behind materializing the graph topology is to empower
the relational database engine with the ability to realize complex graph algorithms. Thus, \ourSys{} helps fill the gap between
the relational model and the massive body of research that assumes a graph model. Listing~\ref{lst:createGraphViewExample} shows how a graph view is created in \ourSys{} from the relational sources of Figure~\ref{Fig:relationalModel}, which is detailed in Section~\ref{sec:CreateGraphView}.

Once a graph view is defined, \ourSys{} allows the user to write pure graph queries, pure relational queries, or queries that mix both graph 
and 
relational operations. \ourSys{}'s query engine views the relational data in either the relational model or the graph model according to the incoming query. In particular, the graph clauses in a query 
are mapped to graph operators in the query execution pipeline, where a graph operator accepts only graph representations as input. \ourSys{} allows the graph operators and the relational operators to co-exist in the same QEP, where the operator type determines the data model of viewing the data (i.e., graph views for the graph model, and relations for the relational model). Section~\ref{sec:CreateGraphView} explains how a graph view is defined from existing relational sources, while Sections~\ref{sec:PathsConstruct} and~\ref{sec:QueryProcessing} demonstrate how to express and evaluate graph-relational queries, respectively.

\section{Graphs as Database Objects}
\label{sec:GraphViews}

As users can create tables in relational databases, they can also create materialized graph views in \ourSys{} as database objects. As we demonstrate in Section~\ref{sec:PathsConstruct},
users can reference the defined graphs using declarative SQL statements. A graph view is created once as a singleton object, and can be referenced by multiple users and queries. 
In Section~\ref{sec:CreateGraphView}, we highlight how graph views can be defined declaratively in \ourSys{}. Section~\ref{sec:Topology} illustrates how the topology of a graph in \ourSys{} is decoupled from the graph data, and how they can be inter-linked. Because dynamic graphs are pervasive in many 
applications, the support for graph updates is addressed in Section~\ref{sec:GraphUpdates}.

\subsection{Creating Graph Views}
\label{sec:CreateGraphView}

\ourSys{} has a declarative \textit{Create Graph View} statement to create graph views initialized from relational data. The statement has 
four
main objectives: (1)~Identifying the name of the graph view to create, 
(2)~Identifying and extracting the graph's set of vertexes from the underlying relational sources, (3)~Identifying and extracting the graph's set of 
edges
from the underlying relational sources, and (4)~Materializing a native graph data structure in memory that reflects the graph topology based on adjacency-list structures. 
Notice that graph traversal operations can be performed efficiently over this native graph representation and is linked back to the corresponding relational data tuples that describe it.
Notice further that the relational source can either be a table or a materialized relational-view because the graph data attributes for the edges and/or the vertexes can be constructed from multiple data sources.

\begin{figure}[t]
\centering
\includegraphics[width=2.0in]
{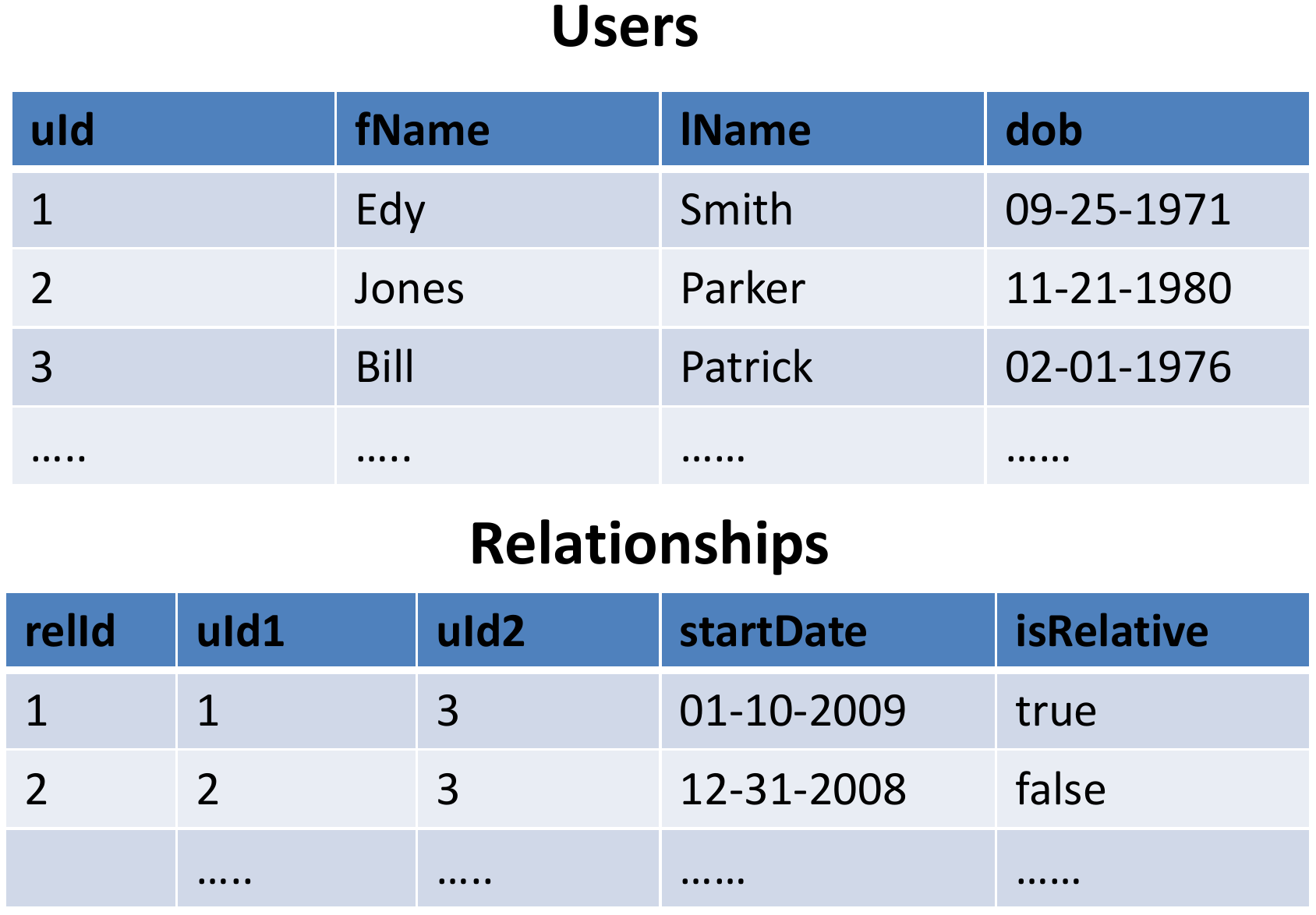}
\caption{A sample social-network in the relational model.}
\label{Fig:relationalModel}
\end{figure}

\begin{figure}[t]
\centering
\includegraphics[width=2.0in]
{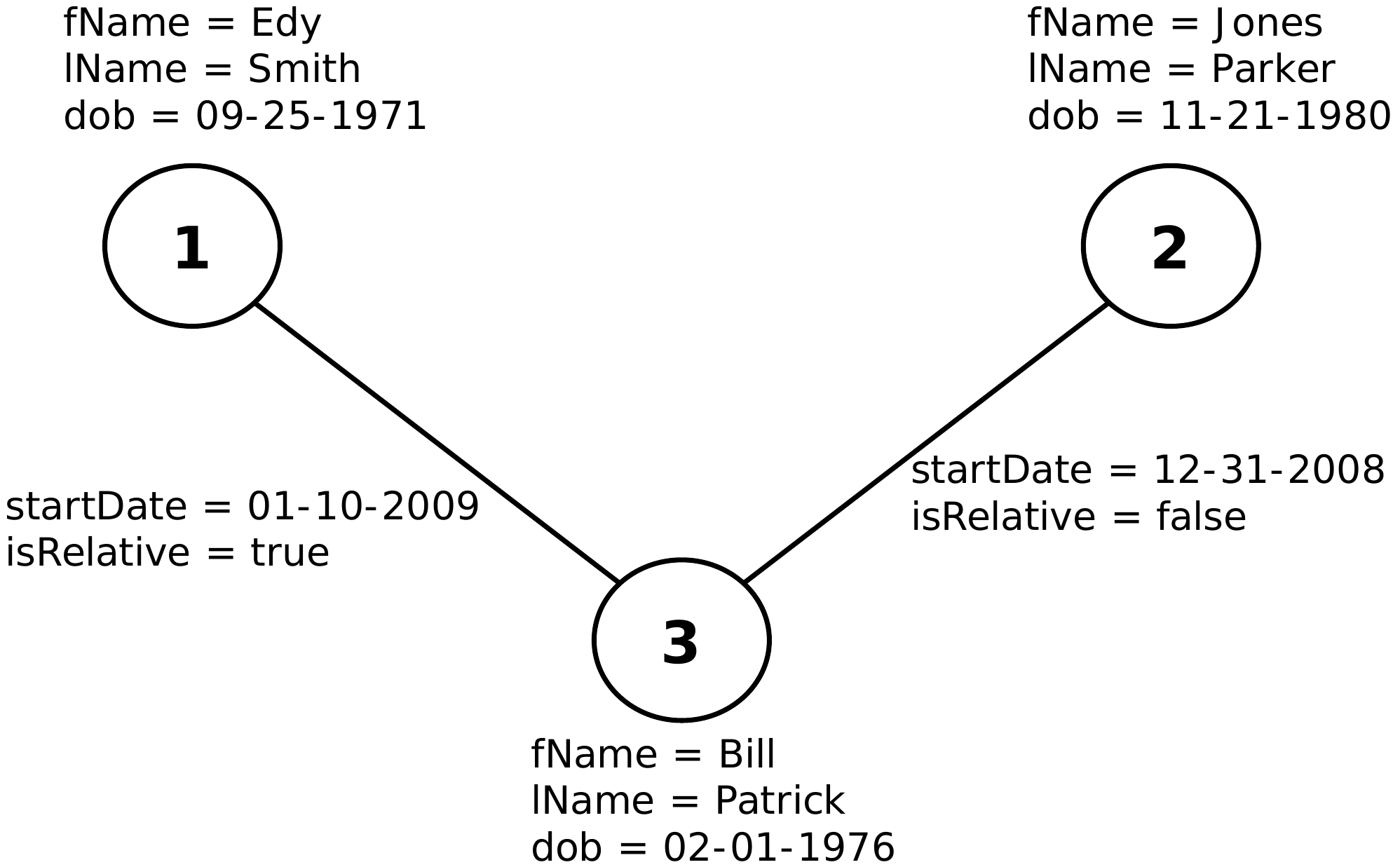}
\caption{A sample social-network in the graph model.}
\label{Fig:graphModel}
\end{figure}

Figure~\ref{Fig:relationalModel} illustrates how a graph view is created in \ourSys{}. 
Assume that the data of a social network is stored in the relational tables as in the figure. 
Tables \textit{Users} and \textit{Relationships} represent the vertexes and the edges of the social network, respectively.
Each vertex or edge has an identifier in the relational tables, which is the only required attribute when creating graph views in \ourSys{}. To illustrate, consider Listing~\ref{lst:createGraphViewExample}, where the relational sources of Figure~\ref{Fig:relationalModel} are used to create a graph view of the social network
of
Figure~\ref{Fig:graphModel}, namely the \textit{SocialNetwork} graph view. A vertex in the \textit{SocialNetwork} graph has its Id from Users.uId and has the two attributes lName and birthdate that get their values from Users.lName and Users.dob, respectively. Similarly, Table \textit{Relationships} defines the edges of the \textit{SocialNetwork} graph, where the edge Id comes from Relationships.relId, and the two edge attributes $sDate$, $relative$ refer to Attributes $startDate$, $isRelative$ of Table \textit{Relationships}, respectively.
For the graph view defined by the \textit{Create Graph View} statement, if the set of vertexes is V, and the set of edges is E,
then, 
the endpoints of an edge in E are constrained to be included in V. \ourSys{} maintains this constraint by ignoring any edge in the \textit{edges relational-source} with endpoints not included in the \textit{vertexes relational-source} of a graph view. 
Other constraint semantics are possible. We pick the one above for simplicity.

\begin{lstlisting}[
      language=graphRelationalLang,
      showspaces=false,
      basicstyle=\ttfamily,
      numbers=none,
      caption={A Social Network Graph View Example},label={lst:createGraphViewExample}]
CREATE UNDIRECTED GRAPH VIEW SocialNetwork
VERTEXES(ID = uId, lstName = lName, birthdate = dob) FROM Users
EDGES (ID = relId, FROM = uId1, TO = uId2, sDate = startDate, relative = isRelative) FROM Relationships
\end{lstlisting}

\subsection{Decoupling the Graph Topology and the Graph Data}
\label{sec:Topology}

\begin{figure}[t]
\centering
\includegraphics[width=2.30in]
{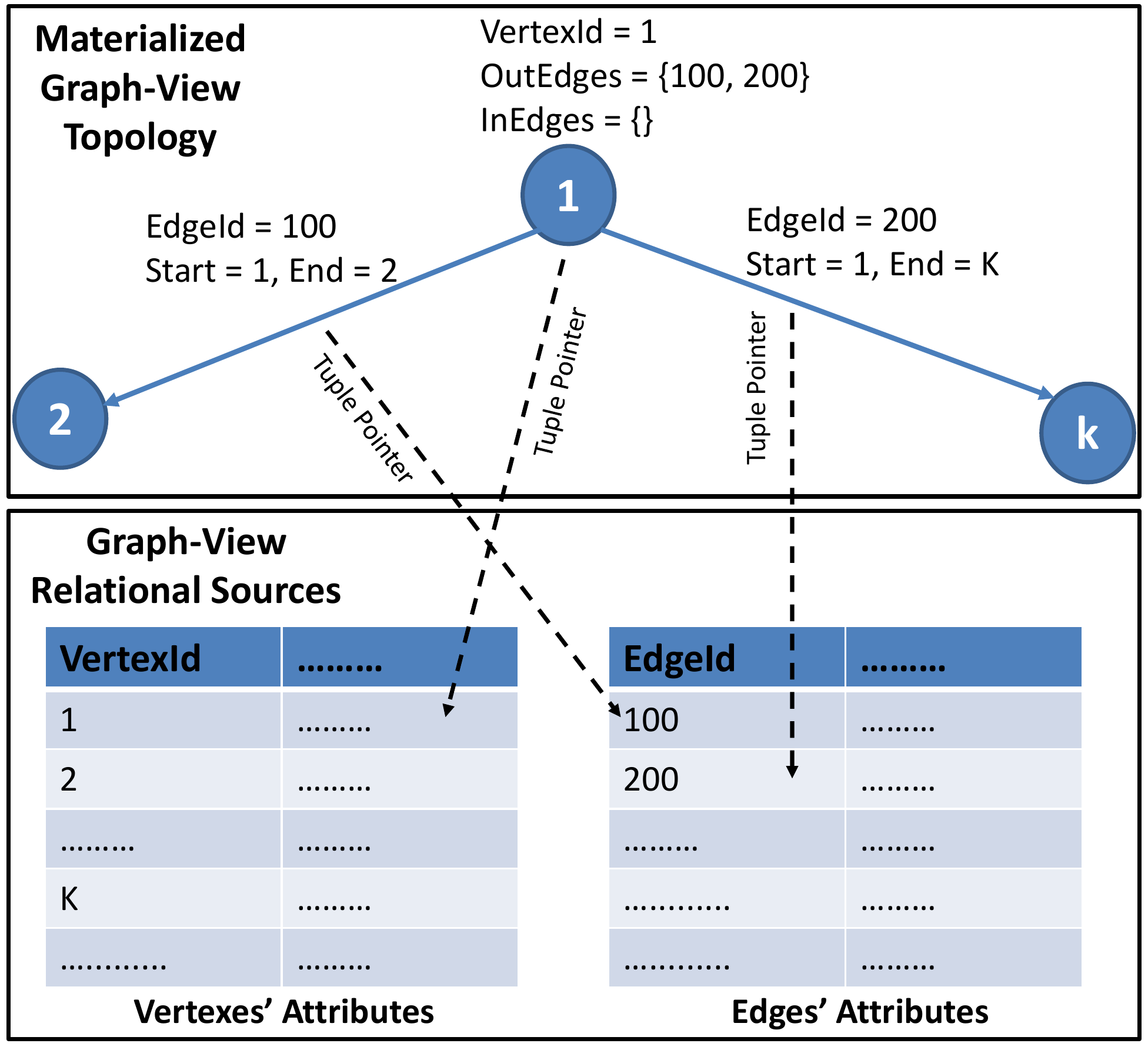}
\caption{A graph view materializes the topology and holds pointers to the relational data of both the vertexes and the edges.}
\label{Fig:graphViewMemory}
\end{figure}

The \textit{Create Graph View} statement updates the system catalog of \ourSys{} to store the definition of the graph view.
Moreover, creating a graph view results in the materialization of the graph topology as a native graph structure in the main-memory 
managed by \ourSys{} (as a singleton object that multiple users and queries can reference). However, the attributes of the vertexes and the edges stored in the relational sources are not replicated in the native graph structure, and 
main-memory 
tuple pointers are used to link the graph topology to the relational sources. To illustrate, Figure~\ref{Fig:graphViewMemory} 
demonstrates 
how the graph topology is separated from the graph data (i.e., the relational attributes of the vertexes and the edges). As in Figure~\ref{Fig:graphViewMemory}, each vertex or edge has a 
main-memory 
tuple pointer that points to the corresponding relational tuple storing the attributes of this vertex or edge.
Notice that the design of \ourSys{} allows a vertex or edge in a graph topology to store multiple tuple-pointers if the relational sources are vertically partitioned (e.g., to support semistructured RDF data ,
where not all the vertexes or edges share the same set of attributes). However, in this paper, we assume a single tuple pointer per vertex or edge because our focus is on exploring the benefits of empowering an RDBMS with native graph-processing.

The graph topology follows the graph model, where 
the topology
is represented physically as a graph data-structure. The key idea behind this native graph representation is to allow for the efficient execution of graph traversals, where relational joins can be mitigated when traversing a graph. The reason is that 
materializing the topology
of a graph view can be 
thought of 
as a traversal index, where each vertex, say $V$, is associated with the identifiers of both the outgoing edges and the incoming edges of 
$V$. Given a graph view, say $GV$, its topology can be constructed using a single pass over the relational sources defining the vertexes and the edges of $GV$.

Notice that there is a bi-directional linkage between the graph topology and the graph's corresponding relational data. To illustrate, let $T$ be a relational tuple containing the attribute values of Vertex~$V$. Using the VertexId attribute of $T$, \ourSys{} can locate Vertex~$V$ in the graph representation in $O(1)$ time using the hash map of the native graph structure. Also, using the tuple pointer associated with Vertex~$V$ in the graph data-structure, Tuple~$T$ can be located in $O(1)$ time.
The benefit of separating the graph topology from the graph data is two-fold. First, the size of the graph view is not affected by the size of the graph data that can be very large in some cases. Second, the attributes of the vertexes and the edges in the relational sources can be easily updated without affecting the native graph representation.\\

\subsection{Graph Updates}
\label{sec:GraphUpdates}

\ourSys{} supports graph-view updates that affect the topology or the attributes stored in the relational sources. The topology is affected only when vertexes or edges are added or deleted.

\subsubsection{Graph-Data Updates}
Updating the attribute data of an edge or vertex is straightforward as the attributes are stored in relations outside the native graph representation. Hence, these relational attributes can be updated directly.
However, updating the \textit{VertexId} and the \textit{EdgeId} attributes 
need 
special handling because these attributes are used 
for navigating
from the relational store to the native graph structure (e.g., to probe path-traversal operators in a QEP as in Section~\ref{sec:QueryProcessing}).
Although updating the identifiers are not common,
\ourSys{} maintains the consistency of the identifiers in the graph representation when updating their corresponding attributes in the relational sources. Also, \ourSys{} maintains the referential integrity of the \textit{edges relational-source} when updating a vertex identifier in the \textit{vertexes relational-source}.

\subsubsection{Graph-Topology Updates}

\ourSys{} allows topological updates when the relational sources are either relational tables or a relational views selecting from a single table. \ourSys{} associates each relational source, say $R$, with the identifiers of the graph views that reference $R$. When inserting a new tuple into $R$, the transaction of the insertion statement updates the graph-view topology as part of the transaction (i.e., adding a new vertex or adding a new edge in the graph representation). Similarly, when deleting a vertex or edge, the deletion statement detects the graph views associated with $R$ and updates the affected graph views accordingly as part of the deletion transaction. For example, if $R$ is an \textit{edges relational-source} for a graph view, say $GV$, the edge in $GV$ corresponding to a deleted tuple is removed from $GV$.

\section{The PATHS Query Construct}
\label{sec:PathsConstruct}
As graph traversal queries form a massive body of graph queries (e.g., reachability and shortest path queries~\cite {Dijkstra_1959,RelApproach_2012_VLDB,HortonDemo,HortonPlus,SP_AlternativePaths,EffectiveCaching}), \ourSys{} extends the SQL language
to declaratively find paths 
in
graph views. \ourSys{} introduces the \textit{PATHS} construct to query its graph views.
For a graph view, say \textit{GV}, \ourSys{} recognizes \textit{GV.PATHS} in the From clause of a select statement. 
Conceptually, this allows \ourSys{} to traverse and retrieve simple paths from \textit{GV} that satisfy a path 
criteria
(e.g., predicates on the attributes of the edges forming the path). In addition to \textit{GV.PATHS}, \ourSys{} recognizes \textit{GV.VERTEXES}, and \textit{GV.EDGES}, to reference the vertexes, and the edges of \textit{GV}, respectively.
We focus on the \textit{GV.PATHS} construct as the other constructs are straightforward.

\ourSys{} models a path as an ordered list of edges, where each edge has a start and end vertexes.
The relational predicates of a query can index the edges and the vertexes of a path, say $PS$, as follows:
\begin{myitem}
\item {\bf PS.Edges[StartIndex..EndIndex].EdgeAttribute}:~References an attribute of the edges starting from $StartIndex$ until $EndIndex$. A value of '*' for the $EndVertex$ placeholder indicates that all the edges starting from $StartIndex$ should satisfy the relational predicate.
\item {\bf PS.Vertexes[StartIndex..EndIndex].VertexAttribute}:~References an attribute of the vertexes starting from $StartIndex$ until $EndIndex$.
A value of '*' for the $EndVertex$ placeholder indicates that all the vertexes starting from $StartIndex$ should satisfy the relational predicate.
\item {\bf PS.Edges[ANY].EdgeAttribute}:~References any edge of Path $PS$, and is used, for instance, to find paths of a specific edge property regardless of 
this edge's 
location in the path (i.e., at least one edge satisfies the predicate).
\item {\bf PS.Vertexes[ANY].VertexAttribute}:~References any vertex in Path $PS$.
\end{myitem}
Observe that the aforementioned \textit{EdgeAttribute}, and the \textit{VerterxAttribute} placeholders can refer to any attribute of the edges or the vertexes that 
have been
defined at the time of creating Graph-view $GV$. In addition, each vertex 
in
Path $PS$ has two additional integral attributes, namely \textit{FanIn} and \textit{FanOut}.
Also, Path~$PS$ allows accessing to some path-specific properties, e.g., \textit{PS.StartVertexId} and \textit{PS.Length} refer to the identifier of the start vertex and the length of Path~$PS$, respectively. 
To illustrate how paths can be queried in \ourSys{}, consider Query $Q_p$ in Listing~\ref{lst:selectPath}. The From clause of $Q_p$ specifies that the paths 
are being traversed
from the SocialNetwork graph view,
where the \textit{vertexes relational-source} of the SocialNetwork graph is Relation~$Users$. The query displays the last names of the friends of friends of all the users with Job = 'Lawyer'. Conceptually, $Q_p$ is evaluated by selecting the sub-graph, say $G_{sub}$, containing edges with start dates after '1/1/2000'.
Using Sub-graph $G_{sub}$, \ourSys{} explores paths consisting of two edges that originate from the vertexes corresponding to lawyers in the social network. Notice that Listing~\ref{lst:selectPath} could use \textit{SocialNetwork.VERTEXES} instead of \textit{Users}. However, Listing~\ref{lst:selectPath} uses the \textit{Users} relation to show how relational tables can be joined with the paths of a graph view.
Notice that the details of the extended query language of \ourSys{} are not the main focus of this paper. However, we provide sample code snippets that are relevant to illustrating the evaluation of the graph-relational queries supported by \ourSys{}.

\begin{lstlisting}[
      language=graphRelationalLang,
      showspaces=false,
      basicstyle=\ttfamily,
      numbers=none,
      caption={Friends-of-Friends Path Query \(Q_p\)},label={lst:selectPath}]
SELECT PS.EndVertex.lstName
FROM Users U, SocialNetwork.Paths PS
WHERE U.Job = 'Lawyer' AND PS.StartVertex.Id = U.uId AND PS.Length = 2 AND PS.Edges[0..*].StartDate > '1/1/2000'
\end{lstlisting}

Listing~\ref{lst:reachabilityQuery} presents a reachability query $Q_r$ that queries a protein-interaction network represented by the BioNetwork graph view, and checks if $Protein~X$ interacts directly (i.e., by an edge) or indirectly (i.e., by a path) with $Protein~Y$ through either a covalent or stable interaction types. \textit{PS.PathString} corresponds to the string representation of Path~$PS$. Notice that many paths can exist between the vertexes corresponding to the specified proteins. So, Query~$Q_r$ uses the \textit{LIMIT~1} clause 
because 
retrieving one path is sufficient 
to decide on reachability.

\begin{lstlisting}[
      language=graphRelationalLang,
      showspaces=false,
      basicstyle=\ttfamily,
      numbers=none,
      caption={Reachability Query \(Q_r\)},label={lst:reachabilityQuery}]
SELECT PS.PathString
FROM Proteins Pr1, Proteins Pr2, BioNetwork.Paths PS
WHERE Pr1.Name = 'Protein X' AND Pr2.Name = 'Protein Y' AND PS.StartVertex.Id = Pr1.Id AND PS.EndVertex.Id = Pr2.Id AND PS.Edges[0..*].Type IN ('covalent', 'stable')
LIMIT 1
\end{lstlisting}

In addition to the ability of referencing the attributes of the edges or vertexes forming a path, say $PS$, \ourSys{} allows aggregation functions on the attributes of the vertexes or the edges of $PS$. The aggregate functions on the attributes of paths have the same usage and constraints as those on relational attributes. 
For example, if the edges of $PS$ have an attribute, say $Weight$, a query can compute the sum of the weight values across all the edges of $PS$, i.e., $sum(PS.Edges.Weight)$ can appear in the select-clause of a query to compute the sum of the weights associated with the edges of Path $PS$.

The PATHS construct can also retrieve sub-graphs based on specific patterns (e.g., the topology of the sub-graph, attributes of the vertexes/edges of the subgraph). For instance, finding 
triangular 
structures with specific edge properties, and counting these triangles are important primitives 
for Machine-Learning, 
e.g.,~\cite{Triangle_KDD_09}, where a triangle structure can be viewed as a loop of three edges. Listing~\ref{lst:triangleQuery} presents 
Query~$Q_t$ that counts the number of triangles, where the edges have specific values for their \textit{Label} attribute. Notice the 
use
of the \textit{Path.Length} property, where it is necessary to retrieve only triangles (as the sub-graph of interest has only three edges).

\begin{lstlisting}[
      language=graphRelationalLang,
      showspaces=false,
      basicstyle=\ttfamily,
      numbers=none,
      caption={Subgraph Pattern Query to Find Triangles \(Q_t\)},label={lst:triangleQuery}]
SELECT Count(P)
FROM MLGraph.Paths P Where P.Length = 3 AND P.Edges[0].Label = 'A' AND P.Edges[1].Label = 'B' AND P.Edges[2].Label = 'C' AND P.Edges[2].EndVertex = P.Edges[0].StartVertex
\end{lstlisting}

More interestingly, paths can be joined to query more complex sub-graph patterns. 
Similar to 
relational engines 
that 
can 
perform
self-joins for a relational table, \ourSys{} allows self-joins of the paths of a given graph view.
This is possible as the vertexes and the edges of the paths to join can be referenced by relational join predicates.

\begin{figure}[t]
\centering
\includegraphics[width=1.8in]
{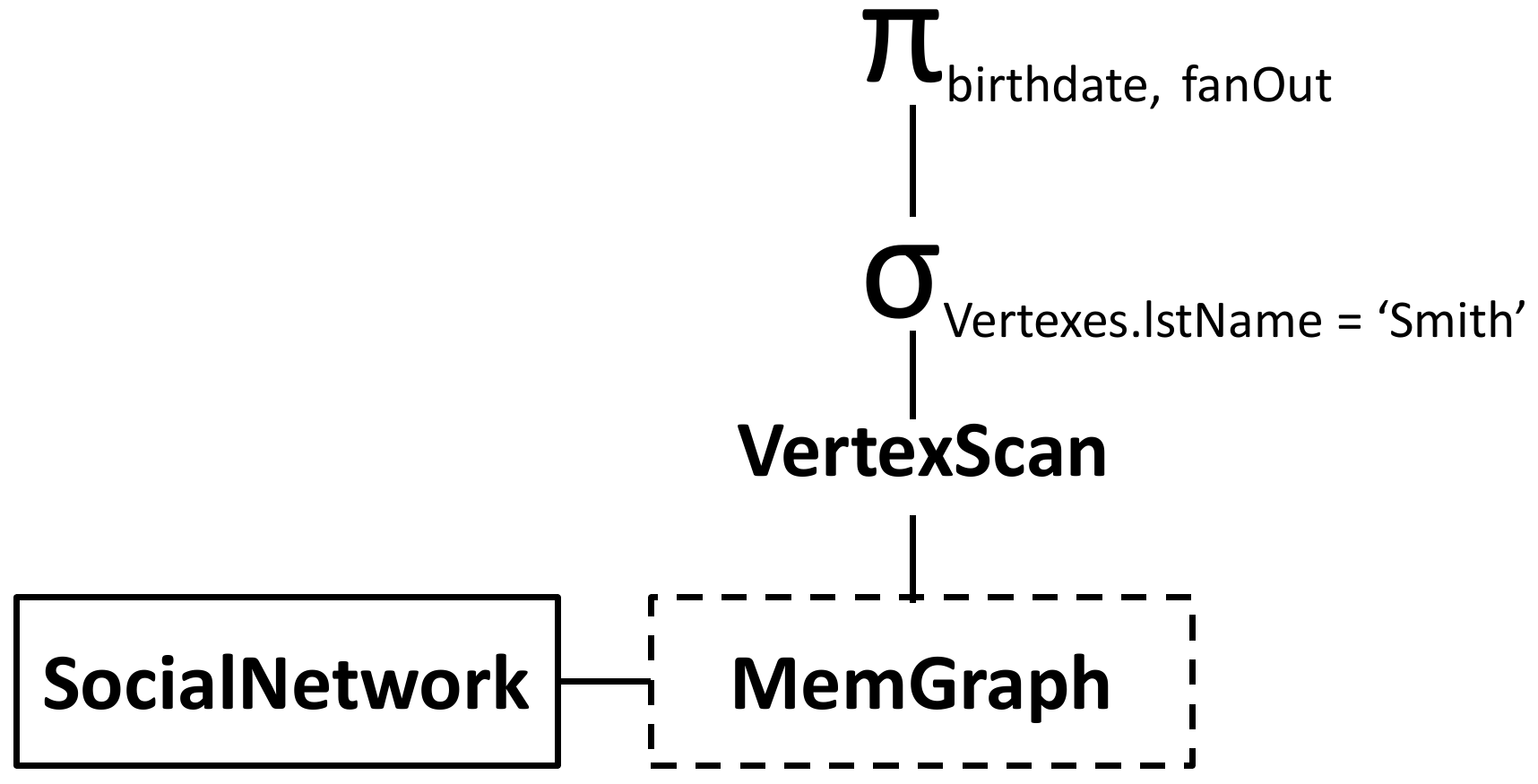}
\caption{QEP for Query $Q_v$.}
\label{Fig:vertexSelect}
\end{figure}




\section{Graph-Relational Query Processing}
\label{sec:QueryProcessing}

In this section, we explain how \ourSys{} evaluates graph-relational queries. Section~\ref{sec:GraphOperators} introduces
the primitive graph operators of \ourSys{}, while Section~\ref{sec:HybridQEP} illustrates how the graph operators 
integrate with typical relational operators in a cross-data-model QEP. Then, Section~\ref{sec:ConceptualEvaluation} discusses the conceptual query evaluation of graph-relational queries in \ourSys{}.

\subsection{Graph Operators}
\label{sec:GraphOperators}

\ourSys{} defines three primitive operators to evaluate the graph constructs of graph-relational queries. In particular, \ourSys{} defines the \textit{VertexScan}, \textit{EdgeScan}, and \textit{PathScan} operators that 
iterate over a graph view's vertexes, edges, and paths, respectively. All the operators in \ourSys{} are lazy operators following the iterator model~\cite{IteratorModel}.

\subsubsection{Vertex Scan and Edge Scan Operators}
\label{sec:VertexScan}

Operators \textit{VertexScan} and \textit{EdgeScan} allow \ourSys{} to iterate over the vertexes and edges of a given graph view, respectively.
For example, the \textit{VertexScan} operator provides an alternative access method for accessing the vertexes of a graph view, where the fan-in and fan-out properties of any vertex can be efficiently retrieved in constant time.
To illustrate, consider Query $Q_v$ in Listing~\ref{lst:selectVertex}. $Q_v$ selects from the set of vertexes of the SocialNetwork graph view, and then performs some relational filtering afterwards. To evaluate $Q_v$, \ourSys{} constructs the query execution pipeline, say $QEP_v$, as in Figure~\ref{Fig:vertexSelect}. Operator \textit{VertexScan} scans the vertexes of the graph defined by the SocialNetwork graph view from the in-memory graph structure (represented as $MemGraph$ in Figure~\ref{Fig:vertexSelect}, that references the singleton graph structure of the graph view). Vertexes with last name 'Smith' are selected and a relational projection operation selects only the birth date and the fan-out properties.

\begin{lstlisting}[
      language=graphRelationalLang,
      showspaces=false,
      basicstyle=\ttfamily,
      numbers=none,
      caption={Vertexes Selection Query},label={lst:selectVertex}]
SELECT VS.birthdate, VS.fanOut
FROM SocialNetwork.Vertexes VS
WHERE VS.lstName = 'Smith'
\end{lstlisting}

\subsubsection{The PathScan Operator}
\label{sec:PathScan}

In \ourSys{}, the \textit{PathScan} operator is responsible for traversing a graph view to construct simple paths 
identified by a graph query. 
\textit{PathScan} is a logical operator that has three physical operators with three corresponding graph-traversal algorithms. All the physical operators explore a traversed vertex only once to avoid loops, i.e., the paths in \ourSys{} are simple paths. In particular, the query optimizer maps a logical \textit{PathScan} operator into $DFScan$, $BFScan$, or $SPScan$, corresponding to depth-first search, breadth-first search, or shortest-path search physical operators, respectively. In this section, we focus on the logical semantics of the path scan operator. We defer the discussion of the physical operators to Section~\ref{sec:QueryOptimization}.

\begin{figure}[t]
\centering
\includegraphics[width=2.3in]
{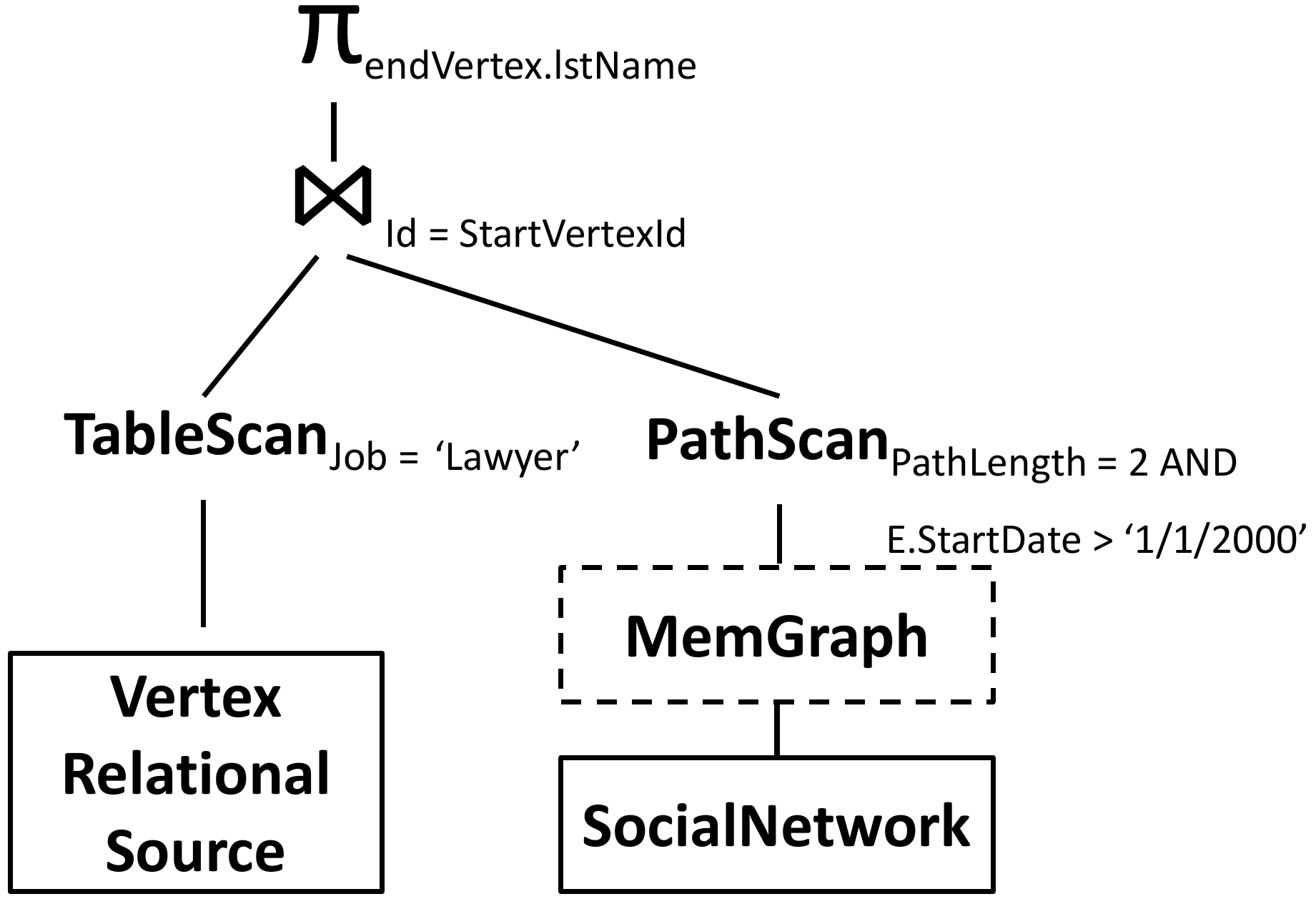}
\caption{\ourSys{} joins a relational table with a graph-view traversal-operator for Query $Q_p$.}
\label{Fig:pathSelect}
\end{figure}

As a logical operation, the paths-discovery process in \ourSys{} starts from a set of start vertexes. These start vertexes are either stated explicitly in the query (e.g., PS.StartVertex.Id = Value) or are generated by some operators during 
query evaluation (e.g., PS.StartVertex.Id = VS.Id as in Listing~\ref{lst:selectPath}). In the latter scenario, the start vertexes selected 
by
some operators (e.g., TableScan, relational sub-query), are used to probe the \textit{PathScan} traversal operator. If the start vertexes of a path selection 
are
not defined, all the vertexes of the corresponding graph view will be used as starting vertexes. To illustrate how paths are explored in \ourSys{}, consider Query $Q_p$ in Listing~\ref{lst:selectPath}. $Q_p$ explicitly states that the path discovery process starts from the vertexes corresponding to lawyers in the social network. Figure~\ref{Fig:pathSelect} 
gives
the query evaluation pipeline $QEP_p$ that evaluates Query $Q_p$, where $MemGraph$ refers to the singleton materialized graph structure of the graph view. In particular, $Q_p$ starts the traversal process
from each qualified vertex. Notice that the qualified vertexes are retrieved using a relational operator (e.g., by a TableScan or IndexScan operators) in Figure~\ref{Fig:pathSelect}. The reason is that using a relational access method with filtering predicates on the \textit{vertexes relational-source} is more efficient than using the tuple pointers in the graph view to filter all the vertexes on the fly.
Because of the seamless integration of the relational and graph models in \ourSys{}, this optimization alternative is feasible.
While traversing the graph view, only the edges with start dates after '1/1/2000' are considered. Also, $QEP_p$ explores paths of length two only (i.e., consisting of two edges) that originate from a given start vertex.
Query evaluation in \ourSys{} follows a lazy-pull evaluation model, where each operator implements the iterator interface~\cite{IteratorModel}. As an effective optimization, \ourSys{} pushes predicates, e.g., path-length predicates, to be considered during the traversal process. This optimization allows \ourSys{} to apply early pruning of paths, and to reduce the size of the intermediate results flowing through the query pipeline. Consequently, the performance of the query evaluation process is boosted w.r.t. the processing time as well as the temporary memory used for the intermediate results.

\subsection{Cross-Model Query-Execution-Pipelines}
\label{sec:HybridQEP}

A query in \ourSys{} can reference relations or relational views with graph views simultaneously.
A pure relational engine has a main structure (i.e., tuple) that is passed between the relational operators in a query evaluation pipeline (QEP). 
\ourSys{} 
allows 
its query engine to view data by two different data models, namely, the relational model and the graph model. 
\ourSys{} allows 
a single QEP to have two main categories of operators that interact seamlessly in a pull-based lazy QEP. The first category contains the relational operators (e.g., select, project, relational join) that can interact directly with relational tables. The second category contains graph operators that can operate on graph views (see Section~\ref{sec:QueryProcessing}). \ourSys{} integrates both categories of operators by allowing a relational operator to operate on the result of a graph operator. In particular, \ourSys{} unifies the interface of the output of both the relational and the graph operators. Specifically, the query engine of \ourSys{} abstracts graph processing by using three data types that extend the $Tuple$ data type, namely the $Vertex$, $Edge$, and $Path$ data types, where each has a schema that depends on 
the queried graph-view, as explained below. 

In \ourSys{}, a
vertex, say $V$, is represented in a QEP by a tuple, say $T$, where each attribute of $V$ becomes an attribute in $T$. For example, a graph vertex in Listing~\ref{lst:createGraphViewExample} is represented by a tuple with attributes: ($uId$, $lstName$, $birthdate$). In addition, Vertex $V$ has the following properties:
\begin{myitem}
\item {\bf FanOut}:~Contains the number of the outgoing edges from $V$.
\item {\bf FanIn}:~Contains the number of the incident edges into $V$.
\end{myitem}
An edge $E$ is represented by a tuple with attributes corresponding to $E$'s attributes in addition to the following attributes:
\begin{myitem}
\item {\bf From}:~Contains the start vertex of Edge $E$.
\item {\bf To}:~Contains the end vertex of Edge $E$.
\end{myitem}

\ourSys{} defines the $Path$ data type, where a path, say $P$, is stored in memory as a sequence of identifiers of the edges 
that form 
$P$. In particular, $P$ is an extended tuple with the following attributes defining its schema:
\begin{myitem}
\item {\bf Length}:~Is the number of edges in $P$.
\item {\bf StartVertex}:~Is the start vertex of $P$.
\item {\bf EndVertex}:~Is the end vertex of $P$.
\item {\bf Vertexes}:~Is the 
list of
vertexes forming $P$.
\item {\bf Edges}:~Is the 
list of
edges forming $P$.
\end{myitem}

\subsection{Conceptual Evaluation of Graph-Relational Queries in \ourSys{}}
\label{sec:ConceptualEvaluation} 

\ourSys{} addresses the impedance mismatch between the graph model and the relational model by unifying the type of the elements that move among the relational and the graph operators within a 
QEP.
To illustrate,
we list below the high-level steps that describe \ourSys{}'s conceptual evaluation of declarative graph-relational queries, i.e., ones that reference relation(s) and graph-view(s):
\begin{myitem}
\item The relational tables and views are joined together using all the relational predicates in the WHERE clause of the query. This step yields a single resultant relation, say $R$.
\item Each graph operator operates on a graph view, say $GV$, using its in-memory singleton graph-structure, say $Mem_{GV}$. In case of using different aliases on the same graph view, each alias is assigned an independent pointer to $Mem_{GV}$.
\item When querying a 
combination 
of relations, relational views, vertexes, edges, or paths, all the graph operators operate only on graph views. Observe that the output of each graph operator is an extended type of the relational $Tuple$ type. Hence, the output of the graph operators can be ingested by the relational operators (e.g., the joins) in the same $QEP$ seamlessly, 
where a relational join outer tuple can be used to probe a graph operator in the inner (e.g., see Figure~\ref{Fig:pathSelect}).
\item The predicates in the WHERE clause of the query that have not been consumed in producing $R$ are used to join $R$ with all the vertexes, edges, and paths referenced by the query.
\item The SELECT list is used to perform the projection operation.
\end{myitem}

\section{Query Optimization}
\label{sec:QueryOptimization}

\ourSys{} optimizes graph-traversal queries with two 
objectives in mind: (1)~pruning undesired paths as early as possible, and (2)~favoring traversal algorithms with less-memory requirements. The second goal is vital as memory should be consumed 
discretely
in an in-memory system. Optimization techniques for early pruning are discussed in Sections~\ref{sec:OptimizationPathLength} and~\ref{sec:OptimizationPushAhead}.
In Section~\ref{sec:OptimizationPhysical}, we address how \ourSys{} selects a traversal algorithm 
that reduces memory consumption as well as how shortest-path traversals are handled.

\subsection{Path Length Inference}
\label{sec:OptimizationPathLength}

The query optimizer of \ourSys{} infers whenever possible, the allowed length of the paths described by the queries. The main objective is to make sure that a path returned from the PathScan operator is unlikely to be rejected by a parent operator (e.g., a join operator) due to a predicate referencing the path length. For instance, if a query has the filter "$PS.Edges[5..*].Att1 = Value$", then PathScan infers that the minimum path length to return is 6 (indexing is zero-based). Hence, PathScan will not return a path of length 5 or less.
Many real-world queries specify the length of the desired paths, e.g., triangle-counting queries~\cite{Triangle_KDD_09} specify a path length of three, the popular friends-of-friends queries restrict a path length to two, and many reachability queries put a cap on the maximum length of the path connecting the queried endpoints. 

For each collection of paths, say $PS$, that is referenced in the $From$-clause, the query optimizer analyzes the predicates referencing the length of $PS$ explicitly (e.g., PS.Length = value), or implicitly (e.g., by analyzing the logical operators as in PS.Edges[5..*].Att1 = X AND PS.Edges[7..9].Att2 = Y), to predict the range of allowed lengths of the paths to return. Then, the inferred path length is 
considered by PathScan while traversing the graph (e.g., an inferred maximum path length of 8 will prune any path of length $\ge 9$).\\

\subsection{Pushing Filters Ahead of Path Scans}
\label{sec:OptimizationPushAhead}

To prune paths early, all the filters related to discovering the paths of a graph view are pushed ahead of the \textit{PathScan} operator. For instance, for a graph view's paths, say $PS$, Predicate "$PS.Edges[0..*].Cost < 10$" 
is 
pushed so that \textit{PathScan} can prune any potential path explored with an edge of cost $\ge 10$. Similarly, predicates 
that refer
to aggregates on a path's attributes will be computed and checked during the \textit{PathScan} evaluation. For example, consider a query, say $Q$, with the predicate "$Sum(PS.Edges.Cost) < 100$". When \textit{PathScan} explores Path~$P$ while evaluating $Q$, \textit{PathScan} will accumulate the cost-attribute of the edges of P during the traversal. If the accumulated cost exceeds 100, $P$ will be 
dropped and will not flow to the operators 
next
in the QEP.\\

\subsection{Logical to Physical Operator Mapping}
\label{sec:OptimizationPhysical}

Recall from Section~\ref{sec:PathScan} that the \textit{PathScan} operator is a logical operator that is mapped into one of three physical traversal operators for execution,
namely, depth-first search, breadth-first search, and shortest-path search based on Dijkstra's algorithm~\cite{Dijkstra_1959}.

The shortest-path physical operator, namely $SPScan$, is very useful in 
top-k shortest path queries. Listing~\ref{lst:selectSPWithHint} 
illustrates 
how the user can instruct the optimizer to use $SPScan$. Given a non-negative numerical edge attribute, $SPScan$ traverses the graph using Dijkstra's algorithm~\cite{Dijkstra_1959}, and returns the next shortest-path as requested (i.e., pulled) by the parent operator in the QEP. $SPScan$ is useful in many applications, e.g., recommendation systems and route discovery, to avoid the costly straightforward plan, i.e., avoid enumerating all paths, then filtering, sorting, and then 
returning the top ones. 

For general graph-traversals where shortest paths are not defined, \ourSys{} can use either a depth-first search (i.e., a $DFScan$ operator), or a breadth-first search (i.e., a $BFScan$ operator). The user can give a query hint to decide on depth-first or breadth-first evaluations. To illustrate how \ourSys{} decides on the physical operator to perform a general graph traversal in the absence of an explicit query-hint, assume that a query, say $Q$, searches for Path $P$ of Length $L$. Assume further that Query $Q$ targets a graph view where the average fan-out is $F$. Following an analysis similar to that in~\cite{DFS_BFS_DBTrans_2013}, a depth-first search can contain on average $F~*~L$ vertexes in its stack data structure. In contrast, a breadth-first search can contain $F^L$ vertexes in its queue data structure. Hence, \ourSys{} uses BFS if $F~<~\sqrt[L-1]{L}$
to optimize for memory.
This optimization is 
applicable
if the path length can be inferred and if we maintain the average fan-out statistic for each graph view in the system catalog. Otherwise, \ourSys{} uses the $DFScan$ operator. Some applications have graphs with large fan-out where the condition above 
is unlikely to hold.
Thus, \ourSys{} 
has $DFScan$ as its default path scan operator.
\ourSys{} has a configuration to store the average fan-out of graph views as a statistics object. If this configuration is enabled, \ourSys{} runs a thread in the backend to compute the average fan-out of the graph views for the first time. As in many commercial database systems, users can update the statistics manually by a command, and the database system can automatically update the statistics in a lazy way under certain conditions (e.g., after a specific number of tuple insertions/deletions).

\begin{lstlisting}[
      language=graphRelationalLang,
      showspaces=false,
      basicstyle=\ttfamily,
      numbers=none,
      caption={Declarative Shortest-Path Query},label={lst:selectSPWithHint}]
SELECT TOP 2 PS
FROM RoadNetwork.Paths PS HINT(SHORTESTPATH(Distance)), RoadNetwork.Vertexes Src, RoadNetwork.Vertexes Dest
WHERE PS.StartVertex.Id = Src.Id AND PS.EndVertex.Id = Dest.Id AND Src.Address = "Address 1" AND Dest.Address = "Address 2"
\end{lstlisting}

\section{Experimental Evaluation}
\label{sec:ExperimentalEvaluation}

In this section, we experimentally evaluate the performance of \ourSys{}, a realization of the proposed \ourApproach{} approach inside a centralized version of VoltDB. 
We compare \ourSys{} to the state of the art of the \rApproach{} approach, namely SQLGraph~\cite{SQLGraph_SIGMOD_2015}, and we compare to Grail~\cite{Grail_CIDR_2015} in Appendix~\ref{sec:ExpShortestPaths}. Although Grail uses a different computational model than \ourSys{}, they both have the common ground of executing queries through an RDBMS. We also compare \ourSys{} to two popular specialized graph systems, Neo4j~\cite{Neo4j} and Titan~\cite{Titan}. The reason 
for
comparing with specialized graph systems is to show that graph-traversal queries can be efficiently handled by \ourSys{}. Notice that from the query-functionality aspect, specialized graph systems are powerful and can answer more queries than those handled by the systems following the \rApproach{}, the \gApproach{}, and the \ourApproach{} approaches.

\noindent
{\bf Mitigating the disk IO cost from the baselines:} As \ourSys{} is an in-memory system, the experiments are designed to mitigate the disk cost of all the baselines we compare to. We implemented SQLGraph and Grail as described in~\cite{SQLGraph_SIGMOD_2015}, and~\cite{Grail_CIDR_2015}, respectively, on top of the in-memory VoltDB system. We configured Titan to use the in-memory storage configuration, and we set Neo4j to run and execute over a RAM disk on Linux.

We consider two important categories of graph queries, namely, traversal-based queries and pattern-matching queries, where the queries can take additional 
filtering predicates. For 
traversal-based queries, we evaluate reachability queries (e.g., Listing~\ref{lst:reachabilityQuery}). We also evaluate shortest-path queries to compare with Grail~\cite{Grail_CIDR_2015} in Appendix~\ref{sec:ExpShortestPaths}. For 
pattern-matching queries, we evaluate the triangle-counting query using filtering predicates on the edges 
while
varying selectivity. The triangle-counting query is a primitive operator in many machine-learning and knowledge-discovery techniques, e.g.,~\cite{Triangle_KDD_09}.
Experiments are conducted on a machine running Linux kernel~3.17.7 on~32 cores of Intel Xeon 2.90~GHz with 384~GB of main-memory.

\subsection{Datasets}
\label{sec:Datasets}
We use real graph datasets that represent four different application domains, namely, road networks, biological networks, authorship networks, and social networks. For the road networks, we use the continental-sized Tiger dataset~\cite{Tiger} that covers the entire U.S., where the edges represent road segments, and the vertexes represent road intersections. For the biological networks, we use the String protein-interaction dataset~\cite{StringDB}, where the vertexes represent proteins, and the edges represent interactions among the proteins. We use the DBLP~\cite{DBLP} dataset for the authorship networks, where the vertexes represent authors, and the edges represent co-authorship relations.
We use the Twitter dataset~\cite{twitter} for the social-network application, where the dataset represents the follower graph of Twitter. The vertexes in Twitter represent users, where an edge from User~A to User~B denotes that User~A follows User~B. Table~\ref{table:Datasets} summarizes the properties of these datasets. \\
\noindent

{\bf Controlling sub-graph selectivity:} We study the effect of selecting a subgraph from an underlying graph before performing a graph operation (e.g., selecting a sub-graph containing 10\% of the edges of the entire graph before executing a shortest-path query or a topological pattern-matching query on the selected sub-graph). 
For each dataset, we vary the selectivity of the queries from 5\% to 50\%.

{\bf Evaluating the effect of graph-views in the \ourApproach{} approach:} To accurately study the performance gains due to the graph-views of the \ourApproach{} w.r.t. the \rApproach approach, we use breadth-firth search instead of depth-first search, and we do not push the predicates ahead of the path scan operator in \ourSys{} for all the reachability-queries experiments.

\begin{table*}
\centering
\begin{tabular}{|l||c|c||c|c|} \hline
Dataset& Number of Vertexes& Number of Edges& Construction Time& Memory Size (GB)\\ \hline
\textbf{Tiger Road Network}&24,412,259&58,698,439&
2.08 Min&0.88\\ \hline
\textbf{DBLP Co-Author Network}&1,007,047&6,592,656&
1.59 Sec&0.09\\ \hline
\textbf{String Protein Network}&1,520,673&348,473,440&
3.81 Min&4.17\\ \hline
\textbf{Twitter Follower Network}&41,652,230&1,468,365,182&
10.87 Min&17.81\\ \hline
\end{tabular}
\caption{The graph views in \ourSys{} are fast to construct with low memory overhead for the datasets of the evaluation.}
\label{table:Datasets}
\end{table*}

\subsection{Unconstrained Reachability Queries}
\label{sec:ExpUnConstrainedReachability}

In this set of experiments, we contrast the performance of \ourSys{} with that of SQLGraph, Neo4j, and Titan, when processing reachability queries without filtering predicates on the graph edges. Given two nodes, say $A$ and $B$, a reachability query returns true if a path exists from Node $A$ to Node $B$. The query-processing time of a reachability query is affected by the path length of the query result. 
The reason is that the increase in the number of edges traversed directly corresponds to the number of relational joins in the \rApproach{} approach (e.g., SQLGraph).

For each dataset in Table~\ref{table:Datasets}, we generate random reachability queries with different path lengths that make the query endpoints connected. We vary the path length from 2 to 20. For each path length, say $l$, we generate $10,000$ random queries, say $Q_l$. We run $Q_l$ and measure the average query-processing time using \ourSys{}, SQLGraph, Neo4j, and Titan.

\begin{figure*}[ht]

\centering
  \subfigure[DBLP Dataset]{\includegraphics[width=1.69in]{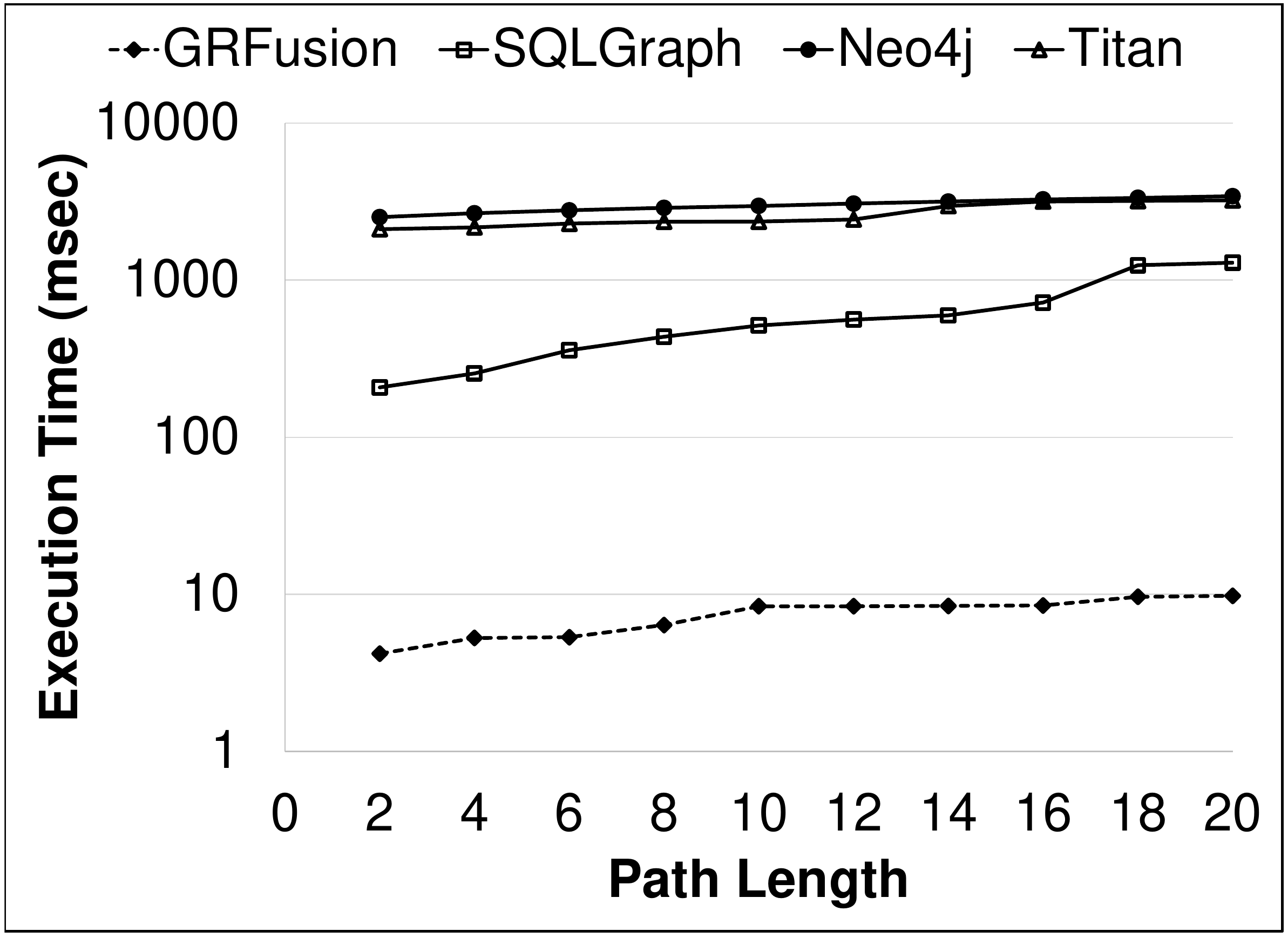}
	\label{Fig:Q01_DBLP}}
 \subfigure[Tiger Dataset]
 {\includegraphics[width=1.69in]{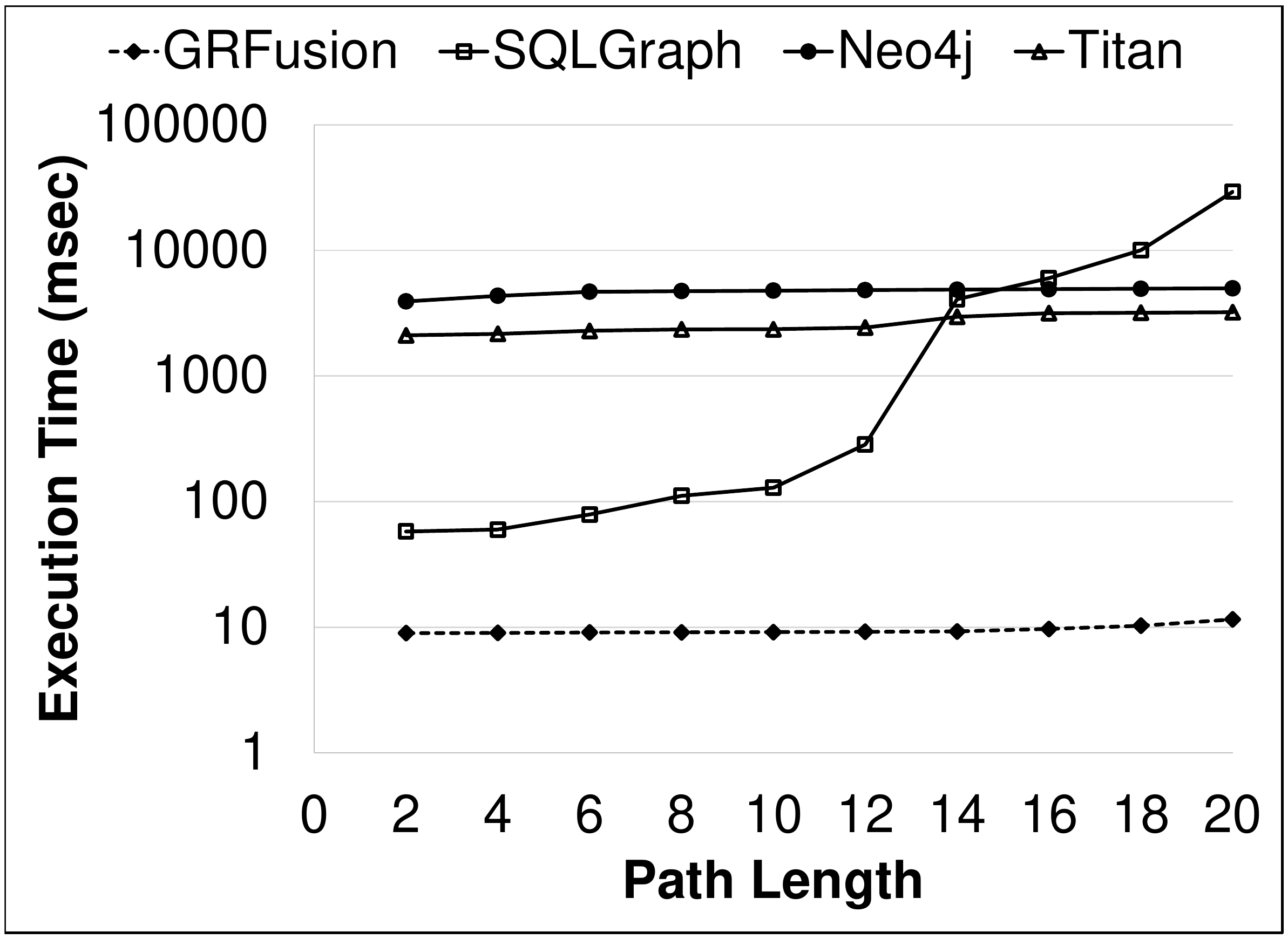} \label{Fig:Q01_Tiger}}
	\subfigure[String Dataset]
	{\includegraphics[width=1.69in]{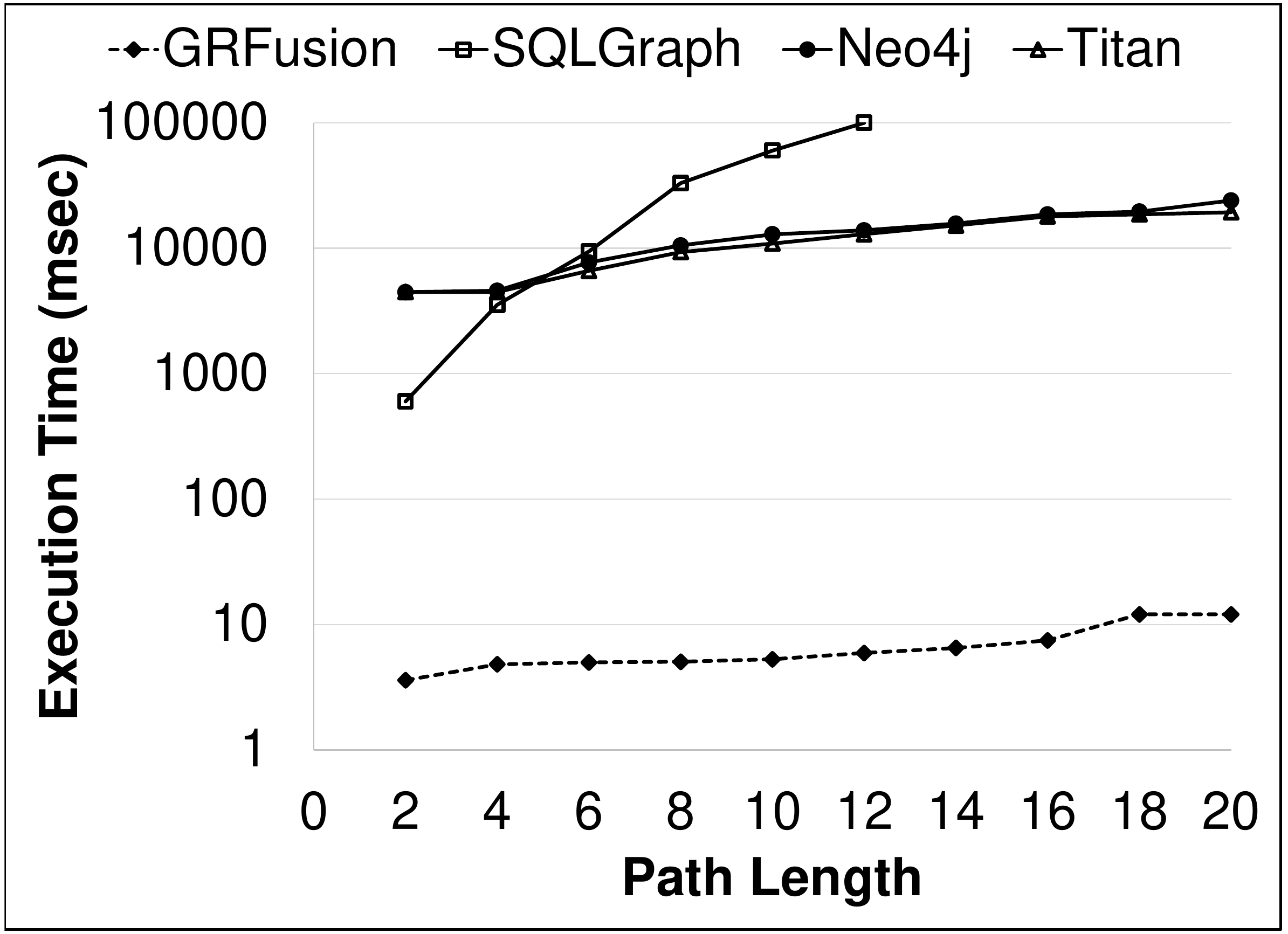} \label{Fig:Q01_String}}
  \subfigure[Twitter Dataset]
	{\includegraphics[width=1.69in]{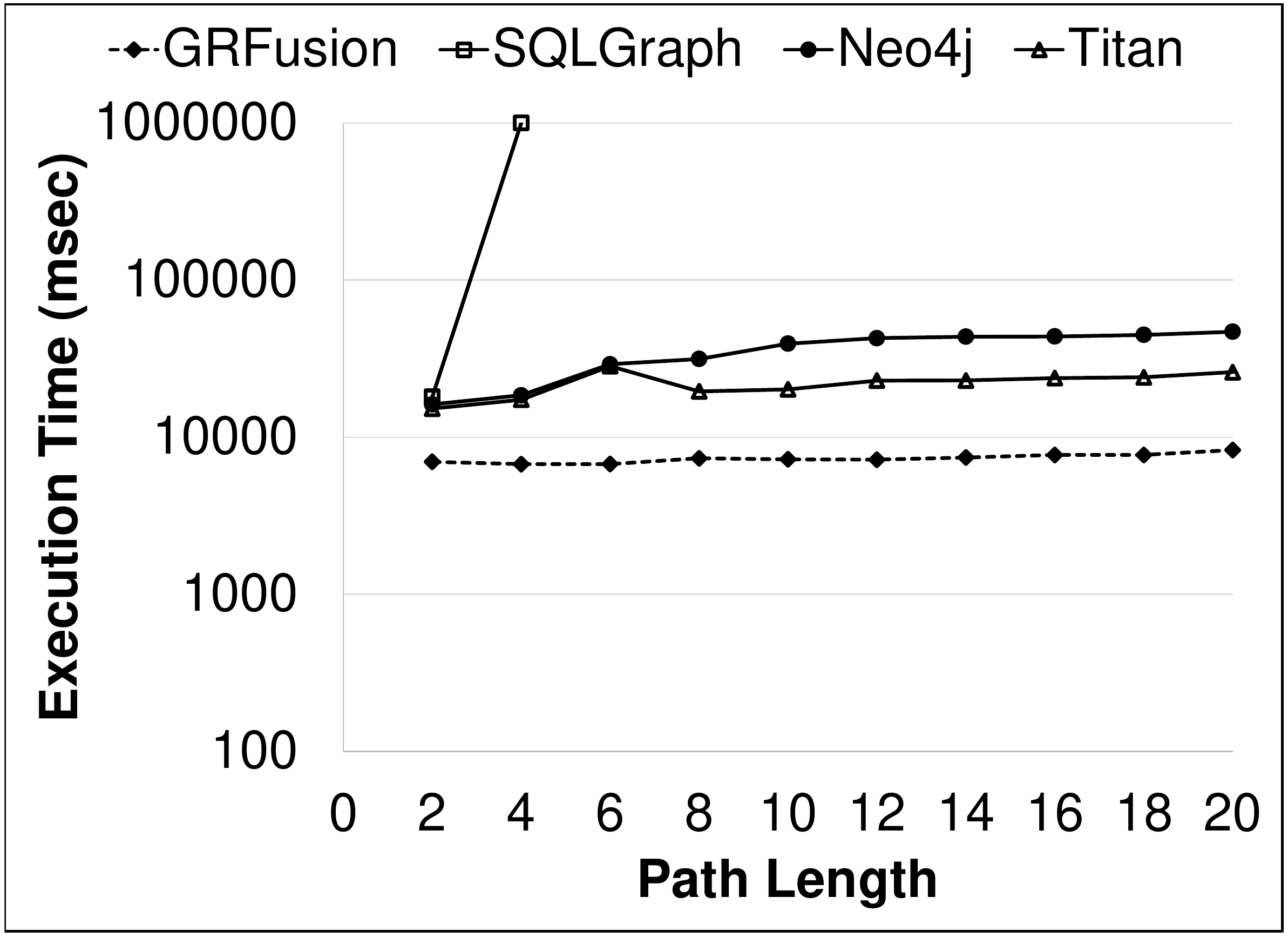} \label{Fig:Q01_Twitter}}
  \caption{\ourSys{} achieves up to 4 orders-of-magnitude speedup in query-time performance for unconstrained reachability queries.}
	\label{Fig:Q01_Reachability_UnConstrained}
\end{figure*}

Figure~\ref{Fig:Q01_Reachability_UnConstrained} gives the average query-processing time of running the queries using all four systems, where the x-axis and the y-axis give the path-length of the query answers and the query-processing time in milliseconds, respectively. 
\ourSys{} achieves up to four orders-of-magnitude speedup in query-processing time compared to SQLGraph, where the speedup increases as the graph size increases. For instance, the speedup reaches 599x for the DBLP graph, and 2483x for the larger String graph. The reason is that \ourSys{} uses the compact graph view that captures the graph topology,
where the graph views act as navigational indexes. Hence, \ourSys{} does not perform any relational join on the relational sources to traverse the graphs. In contrast, SQLGraph performs a relational join for each edge traversal during the path discovery process. Consequently, the query-processing time in SQLGraph increases as the path length of the query result increases.
Moreover, the SQLGraph approach may not scale in main-memory RDBMSs when the graph size is very big due to the size of the intermediate results of the relational joins. To illustrate, 
in Figure~\ref{Fig:Q01_Twitter}, in the Twitter dataset, the \rApproach{} represented by SQLGraph does not execute if the query evaluation requires more than four relational joins. The reason is that the intermediate temporary-memory of the join operators exceeds 6~GB, which is 60 times the 100-MB recommended limit in VoltDB. To allow room for 
query-evaluation pipelining to reduce the intermediate results, and to mitigate the limits of the main-memory, we execute the Twitter queries on a popular disk-based commercial RDBMS. The queries on the Twitter graph time-out after 5 hours of execution when the traversal depth of the queries exceeds four. In contrast, the systems following the \gApproach represented by Neo4j and Titan scale for deep graph-traversal queries on large graphs as the overhead of the relational joins does not exist, where a deep graph-traversal query is a query that explores paths of long lengths, i.e., many edges, which corresponds to many joins in the \rApproach. However, \ourSys{} that realizes the proposed \ourApproach approach combines the benefits of both the relational and the graph worlds, where \ourSys{} is able to scale for deep graph-traversal queries with better performance than 
those 
of the native graph systems. Comparing \ourSys{} to the specialized graph databases Neo4j and Titan, \ourSys{} has a query-time speedup that exceeds three orders-of-magnitude for the String graph (see Figure~\ref{Fig:Q01_String}).
We attribute these performance gains of \ourSys{} over the specialized graph databases to implementation factors and not to a fundamental change in the computational model. The reason is that \ourSys{} is based on VoltDB that has a low-overhead concurrency model (e.g., no lock overhead as in the specialized graph databases). Moreover, VoltDB has an optimized memory manager written in C++ that is significantly more efficient than the JAVA memory managers of both Neo4j and Titan.
Theoretically, if we remove all the implementation-specific factors, the performance of \ourSys{} should be comparable to that of the specialized graph systems as both are processing native graph representations.
In Section~\ref{sec:ExpConstrainedReachability}, we present the performance of \ourSys{} when evaluating queries that do not only consult the graph topology, but also the edges' attributes stored in the relational sources.

\subsection{Reachability Queries with Filtering Predicates}
\label{sec:ExpConstrainedReachability}

In this set of experiments, we evaluate the performance of reachability queries in \ourSys{} and compare it to the baselines when the queries are associated with a filtering predicate. To study the effect of sub-graph selectivity (i.e., selecting the sub-graph to perform the query on), we generate reachability queries similar to the ones described in Section~\ref{sec:ExpUnConstrainedReachability} with varying selectivities. We vary the selectivity parameter from 5\% to 50\% using synthesized edge attributes to control the selectivity.
We 
limit the path length of the results of the generated queries to 20 to emphasize the effect of the selectivity of the sub-graph to operate on.

\begin{figure*}[ht]
\centering
  \subfigure[DBLP Dataset]{\includegraphics[width=1.69in]{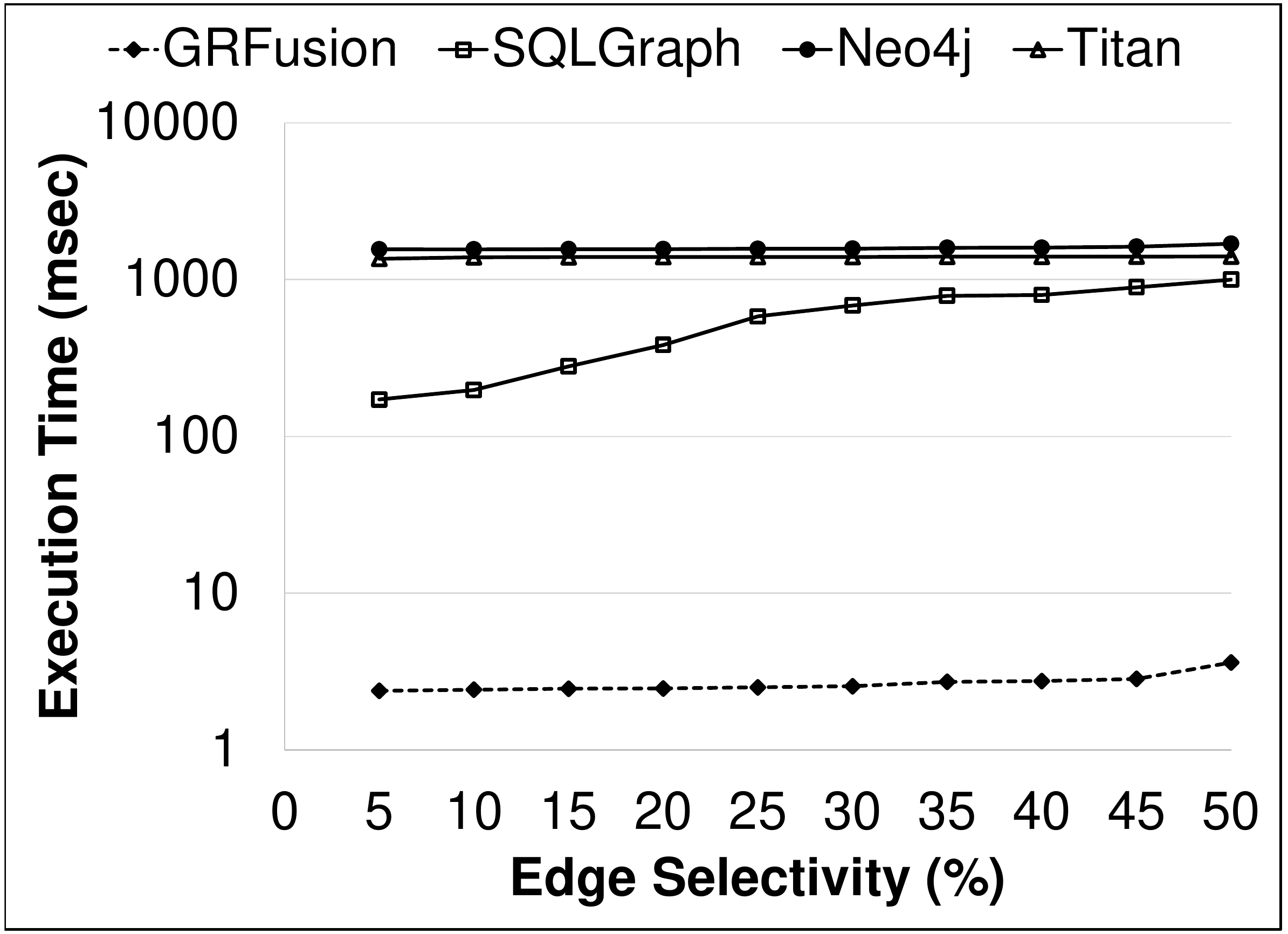}
	\label{Fig:Q02_DBLP}}
 \subfigure[Tiger Dataset]
 {\includegraphics[width=1.69in]{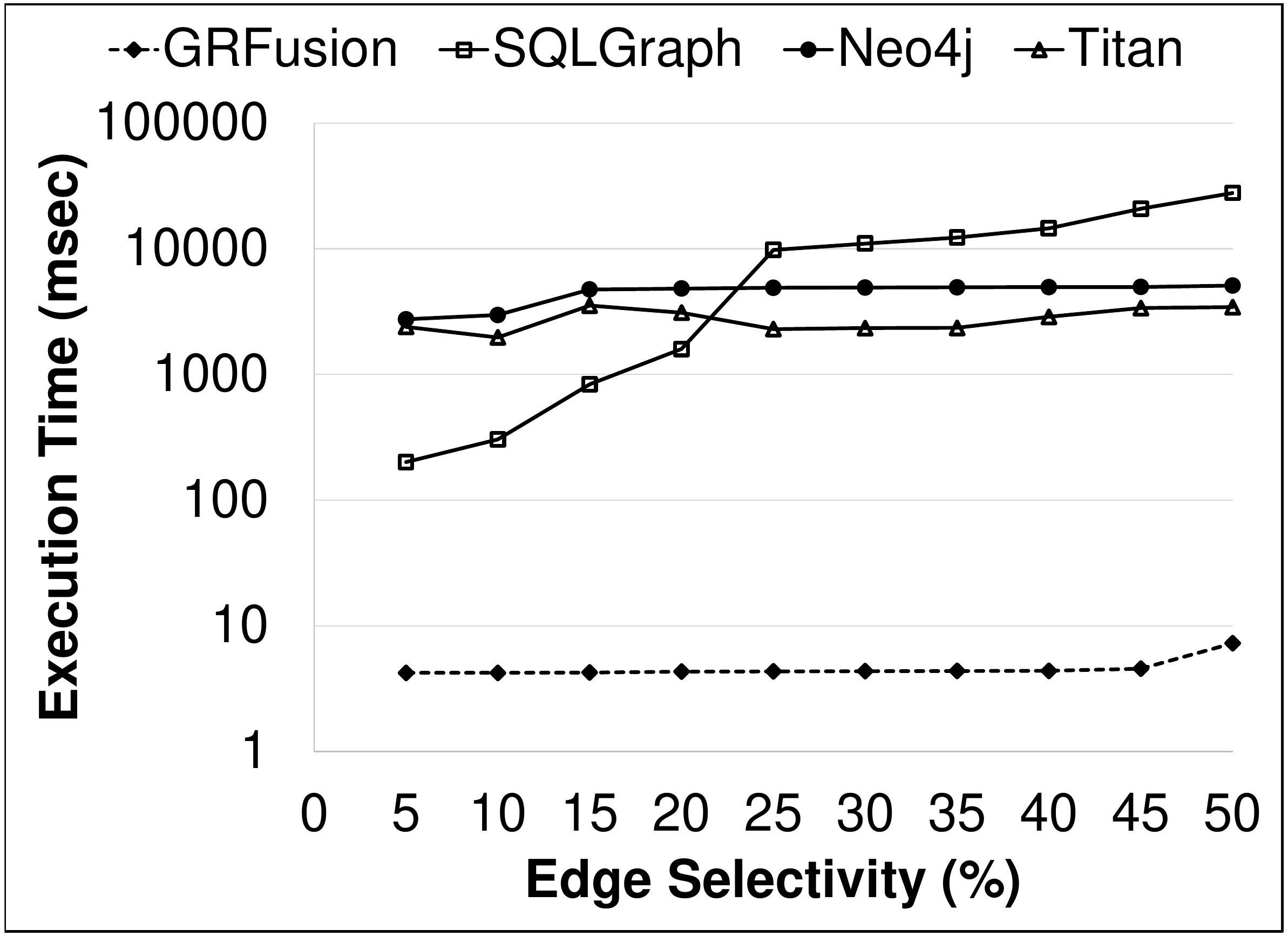} \label{Fig:Q02_Tiger}}
	\subfigure[String Dataset]
	{\includegraphics[width=1.69in]{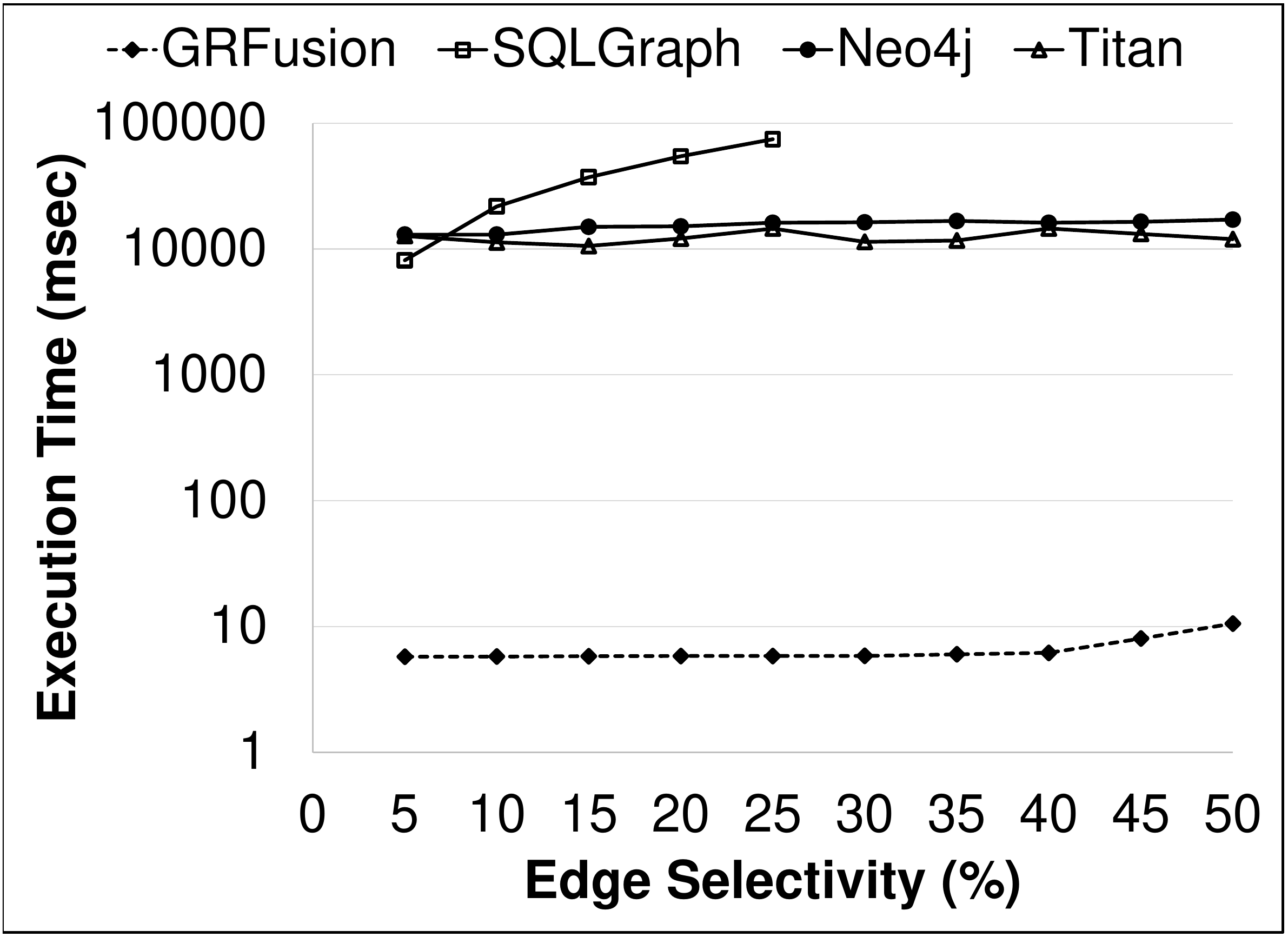} \label{Fig:Q02_String}}
  \subfigure[Twitter Dataset]
	{\includegraphics[width=1.69in]{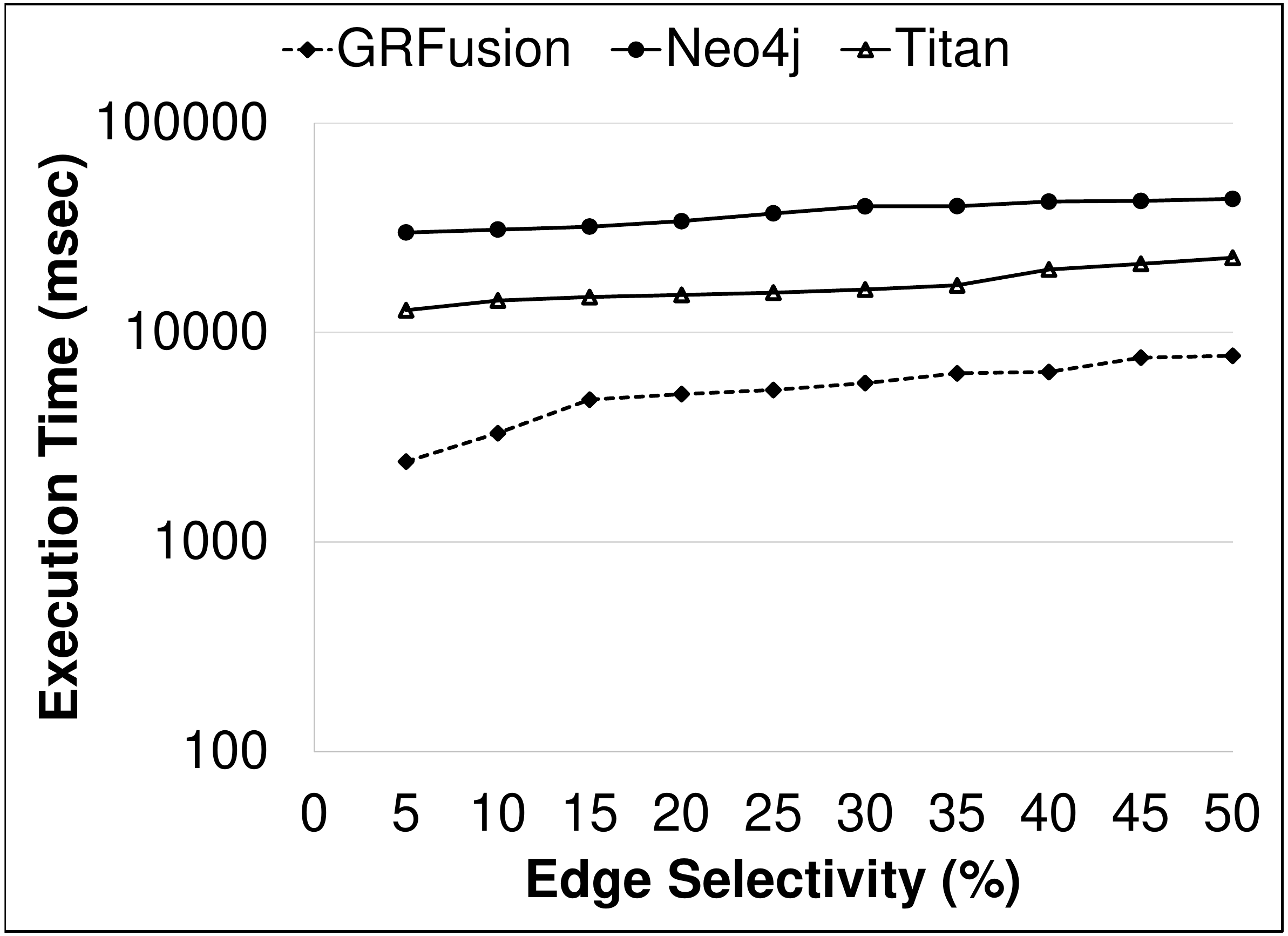} \label{Fig:Q02_Twitter}}
  \caption{\ourSys{} achieves up to 4 orders-of-magnitude speedup in query-time performance for reachability queries with filtering predicates.}
  \label{Fig:Q02_Reachability_Constrained}
\end{figure*}

Figure~\ref{Fig:Q02_Reachability_Constrained} gives the average query-processing time for executing the reachability queries with filtering predicates using all 4 systems and datasets, where the x-axis and the y-axis are the edge-selectivity of the queries, and the query-processing time in milliseconds, respectively. Observe that, for the relatively-small DBLP graph in Figure~\ref{Fig:Q02_DBLP}, SQLGraph outperforms Neo4j and Titan as the relational engine can execute joins and apply filtering predicates efficiently on relations of small cardinalities. 
\ourSys{} outperforms both SQLGraph and the specialized graph engines. There are two main reasons behind \ourSys{}'s performance gains. First, \ourSys{} uses a compact graph data structure to perform the traversal and avoids relational joins completely to explore the underlying graph. Second, \ourSys{} relies on the relational engine to evaluate the filtering predicates on the edges. Recall that \ourSys{} has a direct pointer to an edge's tuple 
that 
is accessed in $O(1)$ time to evaluate the query filtering-predicate using the efficient logic of the relational engine.
Hence, \ourSys{} combines the strengths of both the graph systems and the relational systems to achieve the best-of-both-worlds in terms of performance. However, the efficient evaluation of the filtering predicates and the cost of the relational joins in SQLGraph do not pay off when the size of the relations increase. To illustrate, refer to Figure~\ref{Fig:Q02_Tiger}, where the performance of SQLGraph degrades as more edges are selected. For the String dataset in Figure~\ref{Fig:Q02_String}, SQLGraph exceeds the temporary memory limits of VoltDB after selecting a subgraph of size large than 25\% of the queried graph for the reasons illustrated in Section~\ref{sec:ExpUnConstrainedReachability}. For the largest Twitter dataset, SQLGraph is not able to perform even on a subgraph of a 5\% selectivity. The reason is that the cost of 20 relational joins on the large Twitter table exceeds the temporary-memory limits of VoltDB, and time-out the queries on a commercial disk-based RDBMS after 5 hours of execution. Also, as the number of self-joins increases in the \rApproach{} approach, the relational optimizer may not be able to select the best join algorithm due to inaccurate cardinality estimations of the intermediate results (see~\cite{CardinalityIssue_SIGMOD_2010} for details).

The relational engine is already very efficient in performing filtering predicates. 
This set of experiments demonstrates the power of extending the relational engine with a native graph-core processor that is optimized for graph traversals and that uses efficient memory representation. Figure~\ref{Fig:Q02_Reachability_Constrained} demonstrates the scalability and the efficiency of \ourSys{} in contrast to the baselines in handling graph queries with filtering predicates. Notice that increasing the edge-selectivity factor of the queries has less impact on Neo4j, Titan, and \ourSys{} than on SQLGraph w.r.t. query-processing time. The reason is that 
these 
queries are evaluated on a graph structure by performing the filtering predicates on the fly as the graph is 
being 
traversed. The selectivity affects the query performance 
of
all the approaches.
However, it is more impactful in the case of pure-relational evaluation. For example, in Figure~\ref{Fig:Q02_Tiger}, the processing time of SQLGraph increases by 138x when changing the selectivity from 5\% to 50\%, in contrast to an increase of 1.72x in \ourSys{} on the same setup. 

\subsection{Sub-Graph Pattern Matching}
\label{sec:ExpPatternMatching}

\begin{figure*}[ht]
\centering
  \subfigure[DBLP Dataset]{\includegraphics[width=2.24in]{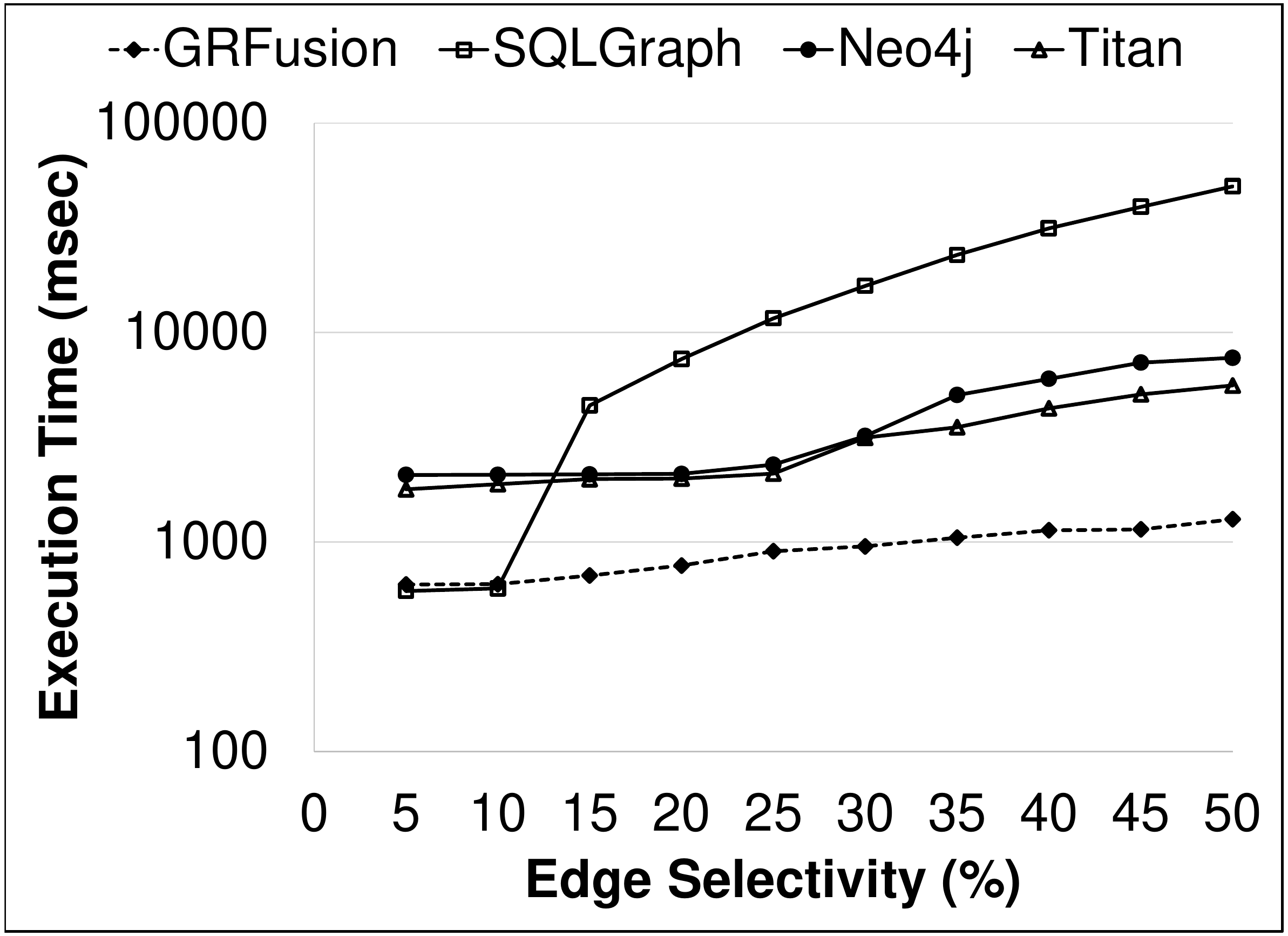}
	\label{Fig:Q12_DBLP}}
  \hspace{0.01in}
 \subfigure[Tiger Dataset]
 {\includegraphics[width=2.24in]{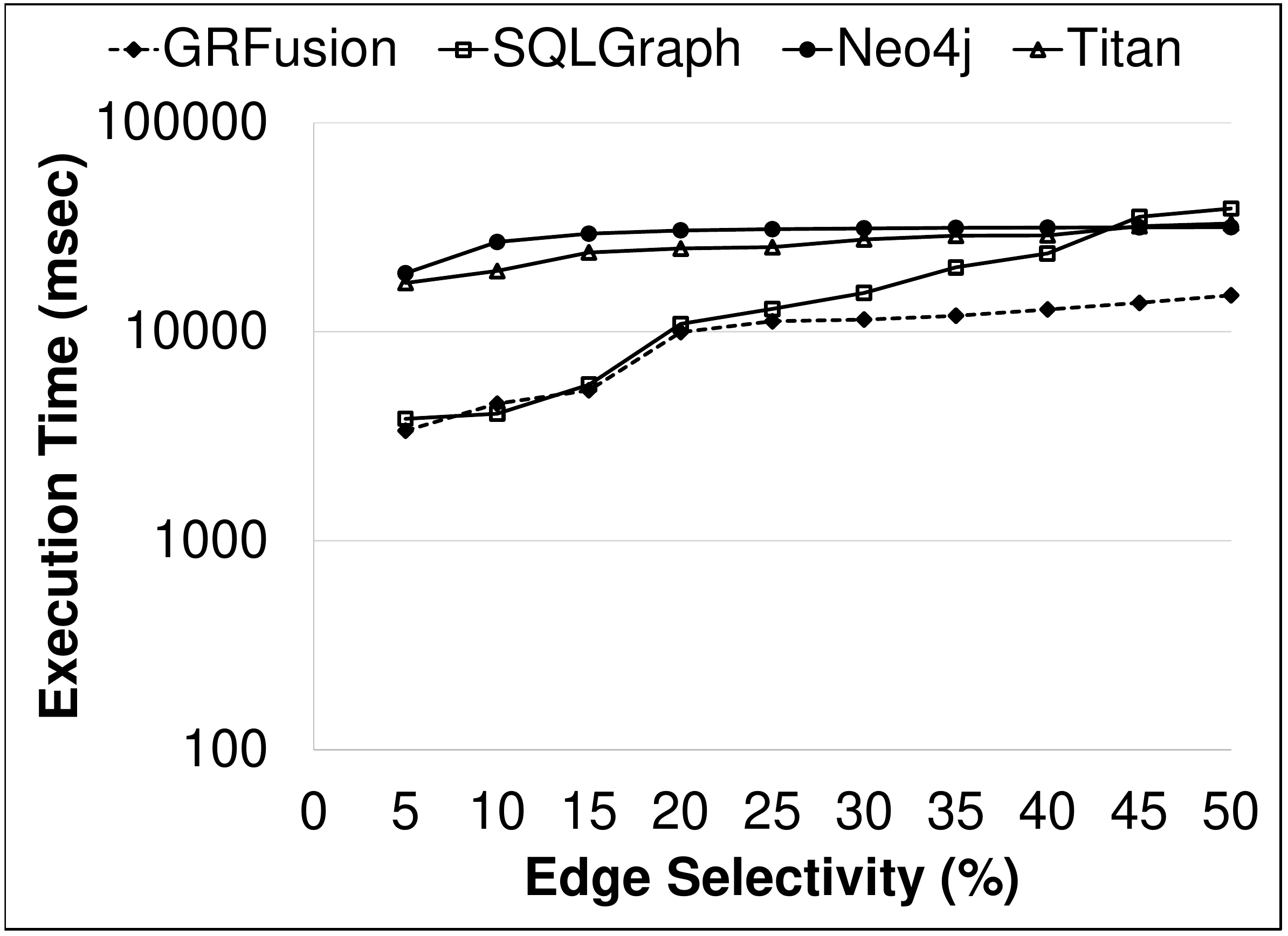} \label{Fig:Q12_Tiger}}
	\hspace{0.01in}
	\subfigure[String Dataset]
	{\includegraphics[width=2.24in]{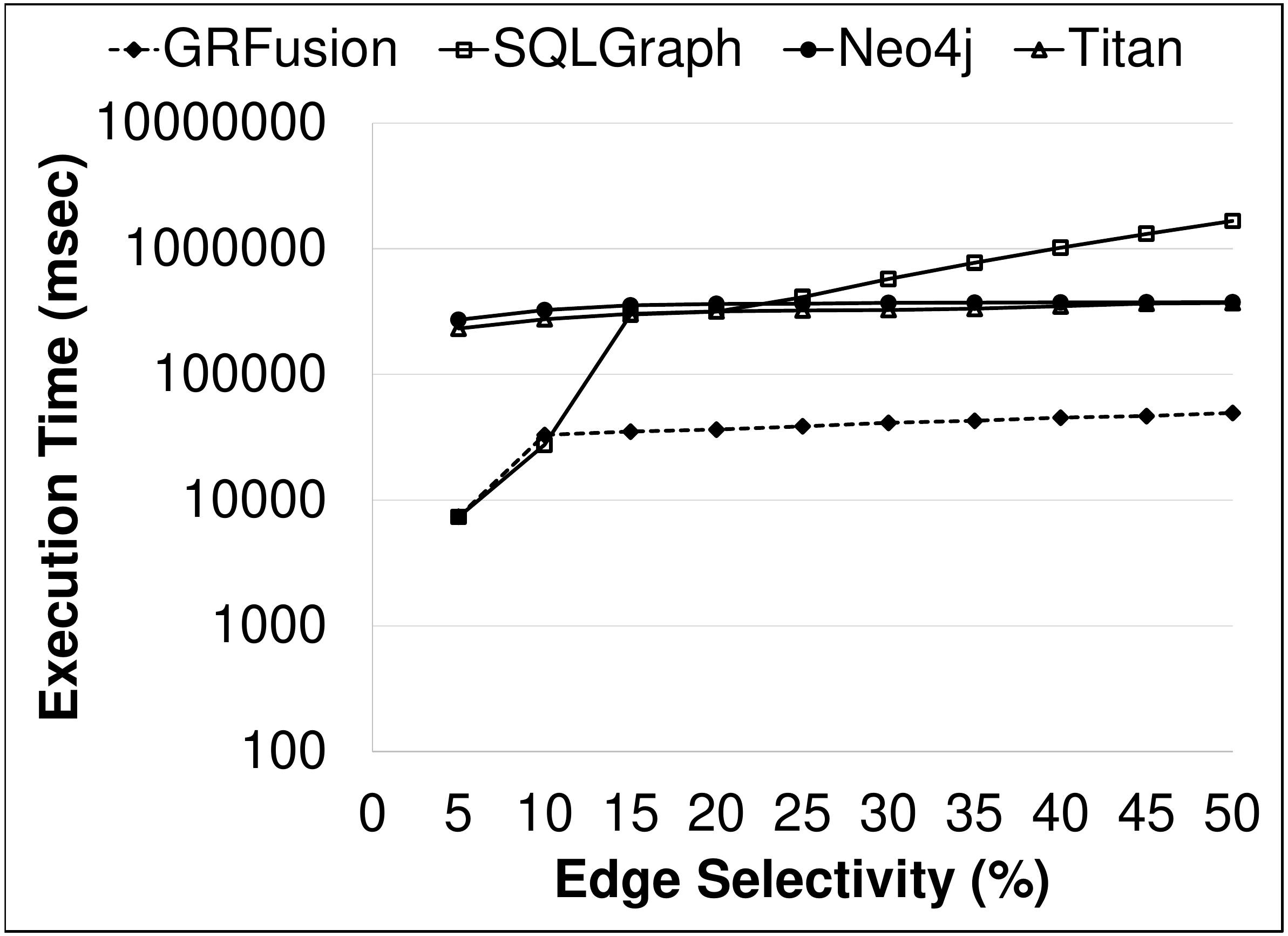} \label{Fig:Q12_String}}
  \caption{\ourSys{} finds all the triangles with filtering predicates with a query-time speedup of one order-of-magnitude w.r.t. SQLGraph and two specialized graph systems.}
	\label{Fig:Q12_TriangleQueries}
\end{figure*}

In this set of experiments, we evaluate the performance of the triangle-counting query.
Given a graph, say $G$, a triangle-counting query, say $Q_{TC}$, finds and counts all the sub-graphs of a triangle pattern (e.g, see Listing~\ref{lst:triangleQuery}). Notice that pure-relational approaches, e.g., 
SQLGraph can scale for this specific pattern query as only two relational joins are needed 
for query evaluation.
This is the reason for choosing
 this pattern query besides its importance as a primitive in many applications~\cite{Triangle_KDD_09}. Figure~\ref{Fig:Q12_TriangleQueries} 
gives
the performance of evaluating triangles queries on the DBLP, Tiger, and String graph datasets, where the 
x-axis and the y-axis
are the edge-selectivity of the 
queries and
the query-processing time in milliseconds, respectively. 

Notice that in Figure~\ref{Fig:Q12_TriangleQueries}, the SQLGraph approach outperforms both Neo4j and Titan when the selected sub-graph size is small, e.g., up to a selectivity of 10\% for the DBLP dataset as in Figure~\ref{Fig:Q12_DBLP}. 
Also, notice 
that SQLGraph is more sensitive to the selectivity parameter than all the other approaches including \ourSys{}. Although only two joins are required by SQLGraph in this type of queries, increasing 
the number of tuples
to join increases the query processing time, which results in better performance by Neo4j and Titan when increasing the selectivity parameter. For instance, Neo4j and Titan are more efficient than SQLGraph for the String dataset in Figure~\ref{Fig:Q12_String} for a selectivity parameter greater than 20\%.

Figure~\ref{Fig:Q12_TriangleQueries} illustrates that \ourSys{} outperforms SQLGraph, Neo4j, and Titan by up to one order of magnitude in query performance. We attribute this performance 
advantage 
by \ourSys{} to the same reasons reported in Section~\ref{sec:ExpUnConstrainedReachability}.

\subsection{The Overhead of Graph Views}
\label{sec:ExpGraphViewsOverhead}

As graph views are materialized in \ourSys{}, we report the construction time as well as the consumed memory space for each dataset. Table~\ref{table:Datasets} illustrates that the construction time ranges from two seconds to 10 minutes according to the size of the graph. The reason is that the construction process passes only once by the vertexes relational-source as well as the edges relational-source. Similarly, Table~\ref{table:Datasets} gives the memory size due to the materialization of the topology of every graph. The consumed memory is of acceptable overhead because only the graph topology is materialized, where each vertex and edge holds pointers to the relational data instead of replicating the relational data inside the graph views. For example, only 0.88 GB is needed to construct a graph view for the continental-sized US road network. Moreover, the overhead of updating the graph views is low. On average, it takes 0.04~milliseconds to add a new edge into an existing graph view, i.e., the total time to insert a tuple in the relational source as well as updating the topology of the corresponding graph view. For both the deletions and insertions of vertexes and edges, \ourSys{} incurs 5\%-11\% additional overhead to the time of manipulating the relational sources. The reason 
for 
this low overhead is that the logic of manipulating the graph views is linear in time w.r.t. the number of affected vertexes or edges as illustrated in Section~\ref{sec:GraphUpdates}. 

\section{Related Work}
\label{sec:RelatedWork}

\textbf{Graphs Integration with Relational Databases}:
There is a plethora of database systems that adopt the graph data model (e.g., Neo4j~\cite{Neo4j} and Titan~\cite{Titan}). These systems have powerful graph querying features. However, it has been shown that for many graph queries, the performance of these systems can be achieved or exceeded by a vanilla relational database~\cite{SQLGraph_SIGMOD_2015,Grail_CIDR_2015}.
For graph-relational queries (which are the main focus of this paper), a graph database is useful if it is feasible to: a)~import all the relational data into the graph database, or b)~develop a custom layer where results from the graph database and the relational database are integrated to form the final results. In contrast, \ourSys{} allows efficient execution of graph-relational queries with neither the overhead of importing data nor the overhead of integrating query results from different systems. Commercial systems, e.g., Oracle~Graph and Aster~\cite{Aster6_VLDB_2014}, follow the architecture of processing graph-relational queries using different run-time systems, where the results are combined at the end. For example, Aster allows defining graph functions that can be referenced in the FROM-Clause of a SQL statement. During query execution, the graph function is extracted and evaluated using a graph runtime system. Eventually, the result from the \textbf{external} graph-runtime is transformed into a relational table that can be integrated with the parent SQL query. In contrast, \ourSys executes the graph operations as well as the relational operations of a query through a cross-data-model QEP without leaving the realm of the RDBMS.

Several graph libraries and systems target graph analytics, e.g., CRAY Graph Engine~\cite{CRAY}, Pregel~\cite{Pregel_SIGMOD_2010} and its open source version Giraph~\cite{Giraph}, GraphLab, GraphFrames~\cite{GraphFrames}.
For graph analytics, it may be acceptable to import data from relational databases for analytical purposes. In contrast, \ourSys{} also serves OLTP scenarios. This is possible as the relational data in \ourSys{} is not deeply copied into the graph views. Moreover, the updates to the relational data that affect the topology of the defined graph views incur little overhead to update the in-memory graph structures in \ourSys{}.

\textbf{Relational Databases with Modified Layers for Graph Processing}: In this category, the internals of a relational database are modified to some extent, but not to a level that executes a graph-relational query through the same QEP as in \ourSys{}.
For example, SAP~HANA~Graph and GRAPHITE~\cite{Graphite_SSDBM_2015} allow graph operations to be directly executed on the relational data in a column-store without data replication.
However, two different runtime components execute the graph-relational queries, i.e., two different query executors, one for the graph portion of the query and the other is for the relational portion.
In contrast, \ourSys{} uses a single runtime to execute graph-relational queries. Thus, query 
optimization
and evaluation methods that allow for the interleaving of relational and graph operators in the same QEP become possible (see Appendix~\ref{sec:ComposingOperators}).
In~\cite{MassiveGraphsInRelationalDatabase_BigData_2013}, an access method is proposed to process graphs stored on disk under certain locality assumptions. In contrast, \ourSys{} is a main-memory system that traverses a graph by realizing a 
light-weight
structure describing the graph topology.  

\textbf{Extracting Graphs from Relational Databases}:
In this category, graphs stored in relational tables are extracted from the database system to be under the control of an independent application. This independent application allows for querying the extracted graphs using graph APIs. Ringo~\cite{Ringo_SIGMOD_15} and GraphGen\cite{GraphGen_Demo_VLDB_15,GraphGen_Paper_SIGMOD_17} are representatives of this approach. Ringo extracts a graph from a relational storage into an in-memory graph structure, where several analytical graph functions are provided that process the in-memory graph. Similarly, GraphGen defines a graph-extraction language~\cite{GraphGen_Paper_SIGMOD_17} to extract latent graphs from relational databases. GraphGen presents APIs to process graph queries targeting the extracted graph, and shows interesting trade-offs for storing graphs in the main-memory. Also, GraphGen allows for importing the extracted graphs into graph databases for advanced graph analytics. In contrast, \ourSys{} processes graphs inside the relational database and does not extract the graphs outside the realm of the database engine. Additionally, \ourSys{} supports dynamic graphs, where online updates are possible. Notice that to support graph-relational queries, e.g., in 
Ringo or GraphGen, the relational part of the query should be processed by the relational database, and the graph operations should be processed by Ringo or GraphGen, where another external layer will be responsible for integrating the graph results and the relational results into the final query result.

\textbf{Encoding Graphs in Relational Databases}: In this line of work (e.g., SQLGraph~\cite{SQLGraph_SIGMOD_2015}, Grail~\cite{Grail_CIDR_2015}), graphs are stored in relational tables with schema optimized for specific graph queries. After storing (or encoding) graphs in a vanilla relational database, a translation layer is designed to translate the supported graph queries into complex SQL statements for the relational database to execute. For instance, SQLGraph~\cite{SQLGraph_SIGMOD_2015} encodes a given property graph in a complex relational schema that is generated based on the graph dataset, and translates specific Gremlin~\cite{Gremlin} queries into SQL statements with multi-join operations for the relational database engine to execute. Although the query performance of this approach is comparable to specialized graph databases for specific queries, these systems do not allow for declarative graph-relational queries. In particular, the schema of the relations storing the graph data may not be suitable for users to query directly and join with other relational data. The reason is that the schema is usually auto-generated based on the input graph for optimization purposes. Also, as SQLGraph is designed for pattern-matching queries that involve few relational joins, it becomes inefficient when handling path queries that involve several relational joins.  

\textbf{Tailored Operators for Specific Graph Operations}: In this category, several research efforts (e.g.,~\cite{Recursive_1988,RecursiveAlgebra_1989_SIGMOD,TransitiveClosure_1991_ICDE}) have been conducted since the 1980s and until recently (e.g.,~\cite{RelApproach_2012_VLDB,EmptyHead_2016_SIGMOD}). However, most of these efforts target specific query types (e.g., transitive closure, shortest paths).
Unlike \ourSys{}, these approaches also do not support a unified/cross-model declarative language to query both graph and relational objects simultaneously.
In~\cite{Recursive_1988,RecursiveAlgebra_1989_SIGMOD}, Relational Algebra is extended with operators to allow for recursive queries. Although the proposed recursive algebra helps execute some graph traversal queries, query execution is not efficient because the graph operators execute over relational tables and not over native graph representations. For instance, several iterations with insertions into temporary tables are needed to keep track of the traversal state. Similarly, Vertica~\cite{Vertica_BigData} presents optimizations for graph-relational queries. 
However, the graph operations 
execute over 
pure relational structures and not on graph representations. Thus, costly relational joins are mandatory in many cases to traverse graphs. In contrast, \ourSys{}'s graph operators process native graph structures in 
main-memory without performing costly joins and without manipulating temporary tables to traverse a graph topology.
Dar et al.~\cite{TransitiveClosure_1991_ICDE} use relational operators repetitively to compute the transitive closure of a graph represented in a predefined relational schema. 
Gao et al.~\cite{RelApproach_2012_VLDB} present specific optimizations to process shortest-path queries over graphs stored in a relational database.
\ourSys{} is more general and can join graph views with relational tables in the same query. Moreover, \ourSys{} addresses the impedance mismatch between the graph model and the relational model. In EmptyHeaded~\cite{EmptyHead_2016_SIGMOD}, graphs in a relational storage are queried using a datalog-like language~\cite{Datalog}. The core idea of EmptyHeaded is to leverage join algorithms with strong theoretical guarantees in addition to using advanced query-compilation techniques.
In contrast, \ourSys{} avoids relational joins completely when traversing the topology of a graph view, where each vertex or edge holds a pointer to the relational tuple describing the vertex's or edge's attributes.

\section{Conclusion}
\label{sec:Conclusion}
We introduce the notions of in-memory materialized graph views, graph operators that seamlessly integrate with relational operators in query evaluation pipelines, memory management, and query optimization techniques for optimizing graph-relational queries. 
\ourSys{} is a realization of the proposed \ourApproach approach inside an open-source in-memory relational DBMS. The key idea behind \ourSys{} is to show the effect of 
extending a relational query processor 
to handle natively and seamlessly graph and relational data
through cross-data-model QEPs.
We introduce the PATH construct, and the extended SQL language of \ourSys{} to 
declaratively express graph-traversal queries with filtering predicates. \ourSys{} constructs in-memory graph structures to capture the graph topology and exploits the relational engine's power in evaluating filtering predicates over the attributes of the edges and vertexes. Consequently, \ourSys{} efficiently handles deep graph-traversal queries without any relational joins to explore the connectives of the vertexes of a graph.
We evaluate \ourSys{} using 
various
graph queries against the state-of-the-art of two main approaches. First, specialized graph engines where \ourSys{} outperforms Neo4j and Titan by up to three orders-of-magnitude in query-time performance. Second, systems following the \rApproach{} approach, where \ourSys{} outperforms SQLGraph and Grail by up to four orders-of-magnitude in query-processing time.

\begin{small}
\bibliographystyle{abbrv}
\bibliography{references}

\begin{thebibliography}{10}

\bibitem{DBLP}
http://dblp.uni-trier.de/xml/.

\bibitem{Giraph}
http://giraph.apache.org/.

\bibitem{twitter}
http://konect.uni-koblenz.de/networks/twitter.

\bibitem{Neo4j}
http://neo4j.com/.

\bibitem{Gremlin}
https://github.com/tinkerpop/gremlin/wiki.

\bibitem{VoltDBGitHub}
https://github.com/voltdb/voltdb/.

\bibitem{StringDB}
http://string-db.org/.

\bibitem{Tiger}
https://www.census.gov/geo/maps-data/data/tiger.html.

\bibitem{OpenStreetMap}
https://www.openstreetmap.org/.

\bibitem{VoltDBCommercial}
https://www.voltdb.com/.

\bibitem{Titan}
http://thinkaurelius.github.io/titan/.

\bibitem{CRAY}
http://www.cray.com/products/analytics/cray-graph-engine.

\bibitem{EmptyHead_2016_SIGMOD}
C.~R. Aberger, S.~Tu, K.~Olukotun, and C.~R{\'e}.
\newblock Emptyheaded: A relational engine for graph processing.
\newblock In {\em Proceedings of the 2016 Intl. Conf. on Management of Data},
  SIGMOD '16, pages 431--446, New York, NY, USA, 2016. ACM.

\bibitem{Recursive_1988}
R.~Agrawal.
\newblock Alpha: An extension of relational algebra to express a class of
  recursive queries.
\newblock {\em IEEE Trans. Softw. Eng.}, 14(7):879--885.

\bibitem{MassiveGraphsInRelationalDatabase_BigData_2013}
R.~Chen.
\newblock Managing massive graphs in relational dbms.
\newblock In {\em IEEE Intl. Conf. on Big Data}, pages 1--8, 2013.

\bibitem{SP_AlternativePaths}
T.~Chondrogiannis, P.~Bouros, J.~Gamper, and U.~Leser.
\newblock Alternative routing: K-shortest paths with limited overlap.
\newblock In {\em Proceedings of the 23rd SIGSPATIAL International Conference
  on Advances in Geographic Information Systems}, SIGSPATIAL '15, pages
  68:1--68:4, 2015.

\bibitem{RecursiveAlgebra_1989_SIGMOD}
L.~S. Colby.
\newblock A recursive algebra and query optimization for nested relations.
\newblock In {\em Proc. of the 1989 ACM SIGMOD Intl. Conf. on Management of
  Data}, pages 273--283.

\bibitem{TransitiveClosure_1991_ICDE}
S.~Dar, R.~Agrawal, and H.~V. Jagadish.
\newblock Optimization of generalized transitive closure queries.
\newblock In {\em Proc. of the Seventh Intl. Conf. on Data Engineering, April
  8-12, 1991}, pages 345--354.

\bibitem{GraphFrames}
A.~Dave, A.~Jindal, L.~E. Li, R.~Xin, J.~Gonzalez, and M.~Zaharia.
\newblock Graphframes: An integrated api for mixing graph and relational
  queries.
\newblock In {\em Proceedings of the Fourth International Workshop on Graph
  Data Management Experiences and Systems}, GRADES '16, 2016.

\bibitem{Hekaton}
C.~Diaconu, C.~Freedman, E.~Ismert, P.-A. Larson, P.~Mittal, R.~Stonecipher,
  N.~Verma, and M.~Zwilling.
\newblock Hekaton: Sql server's memory-optimized oltp engine.
\newblock In {\em SIGMOD}, SIGMOD '13, pages 1243--1254, 2013.

\bibitem{Dijkstra_1959}
E.~W. Dijkstra.
\newblock A note on two problems in connection with graphs.
\newblock {\em Numerical Mathematics}, 1:269--271, 1959.

\bibitem{Grail_CIDR_2015}
J.~Fan, A.~G.~S. Raj, and J.~M. Patel.
\newblock The case against specialized graph analytics engines.
\newblock In {\em {CIDR} 2015, 7th Conf. on Innovative Data Systems Research},
  2015.

\bibitem{RelApproach_2012_VLDB}
J.~Gao, R.~Jin, J.~Zhou, J.~X. Yu, X.~Jiang, and T.~Wang.
\newblock Relational approach for shortest path discovery over large graphs.
\newblock {\em Proc. VLDB Endow.}, 5(4):358--369, Dec. 2011.

\bibitem{CardinalityIssue_SIGMOD_2010}
A.~Ghazal, D.~Seid, A.~Crolotte, and M.~Al-Kateb.
\newblock Adaptive optimizations of recursive queries in teradata.
\newblock In {\em Proceedings of the 2012 ACM SIGMOD International Conference
  on Management of Data}, SIGMOD '12, pages 851--860, 2012.

\bibitem{IteratorModel}
G.~Graefe.
\newblock Query evaluation techniques for large databases.
\newblock {\em ACM Comput. Surv.}, 25(2):73--169, June 1993.

\bibitem{Datalog}
T.~J. Green, S.~S. Huang, B.~T. Loo, and W.~Zhou.
\newblock Datalog and recursive query processing.
\newblock {\em Found. Trends databases}, 5(2):105--195, Nov. 2013.

\bibitem{Vertica_BigData}
A.~Jindal, S.~Madden, M.~Castellanos, and M.~Hsu.
\newblock Graph analytics using vertica relational database.
\newblock In {\em Proceedings of the 2015 IEEE International Conference on Big
  Data (Big Data)}, BIG DATA '15, pages 1191--1200, 2015.

\bibitem{HStore}
R.~Kallman, H.~Kimura, J.~Natkins, A.~Pavlo, A.~Rasin, S.~Zdonik, E.~P.~C.
  Jones, S.~Madden, M.~Stonebraker, Y.~Zhang, J.~Hugg, and D.~J. Abadi.
\newblock H-store: A high-performance, distributed main memory transaction
  processing system.
\newblock {\em Proc. VLDB Endow.}

\bibitem{Mem07_SLACID}
D.~Kernert, F.~K\"{o}hler, and W.~Lehner.
\newblock Slacid - sparse linear algebra in a column-oriented in-memory
  database system.
\newblock In {\em Proceedings of the 26th International Conference on
  Scientific and Statistical Database Management}, SSDBM '14, pages
  11:1--11:12, 2014.

\bibitem{Pregel_SIGMOD_2010}
G.~Malewicz, M.~H. Austern, A.~J. Bik, J.~C. Dehnert, I.~Horn, N.~Leiser, and
  G.~Czajkowski.
\newblock Pregel: A system for large-scale graph processing.
\newblock In {\em SIGMOD}, pages 135--146, 2010.

\bibitem{Mem06_PlacementInCloud}
K.~Molka and G.~Casale.
\newblock Contention-aware workload placement for in-memory databases in cloud
  environments.
\newblock {\em ACM Trans. Model. Perform. Eval. Comput. Syst.}, 2(1):1:1--1:29,
  Sept. 2016.

\bibitem{Mem01_MEMSCALE}
H.~Montaner, F.~Silla, H.~Fr\"{o}ning, and J.~Duato.
\newblock Memscale: In-cluster-memory databases.
\newblock In {\em Proceedings of the 20th ACM International Conference on
  Information and Knowledge Management}, CIKM '11, pages 2569--2572, 2011.

\bibitem{Graphite_SSDBM_2015}
M.~Paradies, W.~Lehner, and C.~Bornh\"{o}vd.
\newblock Graphite: An extensible graph traversal framework for relational
  database management systems.
\newblock In {\em Proc. of the 27th Intl. Conf. on Scientific and Statistical
  Database Management}, SSDBM '15, pages 29:1--29:12.

\bibitem{Ringo_SIGMOD_15}
Y.~Perez, R.~Sosi\v{c}, A.~Banerjee, R.~Puttagunta, M.~Raison, P.~Shah, and
  J.~Leskovec.
\newblock Ringo: Interactive graph analytics on big-memory machines.
\newblock In {\em SIGMOD '15}, 2015.

\bibitem{DFS_BFS_DBTrans_2013}
A.~D. Sarma, H.~Lee, H.~Gonzalez, J.~Madhavan, and A.~Halevy.
\newblock Consistent thinning of large geographical data for map visualization.
\newblock {\em ACM Trans. Database Syst.}, 38(4):22:1--22:35, Dec. 2013.

\bibitem{HortonDemo}
M.~Sarwat, S.~Elnikety, Y.~He, and G.~Kliot.
\newblock Horton: Online query execution engine for large distributed graphs.
\newblock In {\em Proceedings of the 2012 IEEE 28th International Conference on
  Data Engineering}, ICDE '12, pages 1289--1292, 2012.

\bibitem{HortonPlus}
M.~Sarwat, S.~Elnikety, Y.~He, and M.~F. Mokbel.
\newblock Horton+: A distributed system for processing declarative reachability
  queries over partitioned graphs.
\newblock {\em Proc. VLDB Endow.}, 6(14):1918--1929, Sept. 2013.

\bibitem{Mem05_BTree}
A.~Shahvarani and H.-A. Jacobsen.
\newblock A hybrid b+-tree as solution for in-memory indexing on cpu-gpu
  heterogeneous computing platforms.
\newblock In {\em Proceedings of the 2016 International Conference on
  Management of Data}, SIGMOD '16, pages 1523--1538, 2016.

\bibitem{Aster6_VLDB_2014}
D.~Simmen, K.~Schnaitter, J.~Davis, Y.~He, S.~Lohariwala, A.~Mysore, V.~Shenoi,
  M.~Tan, and Y.~Xiao.
\newblock Large-scale graph analytics in aster 6: Bringing context to big data
  discovery.
\newblock {\em Proc. VLDB Endow.}, 7(13):1405--1416, Aug. 2014.

\bibitem{SQLGraph_SIGMOD_2015}
W.~Sun, A.~Fokoue, K.~Srinivas, A.~Kementsietsidis, G.~Hu, and G.~Xie.
\newblock Sqlgraph: An efficient relational-based property graph store.
\newblock In {\em Proc. of the 2015 ACM SIGMOD Intl. Conf. on Management of
  Data}, SIGMOD '15, pages 1887--1901, 2015.

\bibitem{EffectiveCaching}
J.~R. Thomsen, M.~L. Yiu, and C.~S. Jensen.
\newblock Effective caching of shortest paths for location-based services.
\newblock In {\em Proceedings of the 2012 ACM SIGMOD International Conference
  on Management of Data}, SIGMOD '12, pages 313--324, 2012.

\bibitem{Triangle_KDD_09}
C.~E. Tsourakakis, U.~Kang, G.~L. Miller, and C.~Faloutsos.
\newblock Doulion: Counting triangles in massive graphs with a coin.
\newblock In {\em Proceedings of the 15th ACM SIGKDD International Conference
  on Knowledge Discovery and Data Mining}, KDD '09, 2009.

\bibitem{Mem02_Pipelining}
L.~Wang, M.~Zhou, Z.~Zhang, Y.~Yang, A.~Zhou, and D.~Bitton.
\newblock Elastic pipelining in an in-memory database cluster.
\newblock In {\em Proceedings of the 2016 International Conference on
  Management of Data}, SIGMOD '16, pages 1279--1294, New York, NY, USA, 2016.
  ACM.

\bibitem{Mem03_ScalingMulticore}
Z.~Wang, S.~Mu, Y.~Cui, H.~Yi, H.~Chen, and J.~Li.
\newblock Scaling multicore databases via constrained parallel execution.
\newblock In {\em Proceedings of the 2016 International Conference on
  Management of Data}, SIGMOD '16, pages 1643--1658, 2016.

\bibitem{GraphGen_Paper_SIGMOD_17}
K.~Xirogiannopoulos and A.~Deshpande.
\newblock Extracting and analyzing hidden graphs from relational databases [to
  appear].
\newblock In {\em SIGMOD '17}.

\bibitem{GraphGen_Demo_VLDB_15}
K.~Xirogiannopoulos, U.~Khurana, and A.~Deshpande.
\newblock Graphgen: Exploring interesting graphs in relational data.
\newblock {\em Proc. VLDB Endow.}, 8(12):2032--2035, Aug. 2015.

\bibitem{Mem04_Logging}
C.~Yao, D.~Agrawal, G.~Chen, B.~C. Ooi, and S.~Wu.
\newblock Adaptive logging: Optimizing logging and recovery costs in
  distributed in-memory databases.
\newblock In {\em Proceedings of the 2016 International Conference on
  Management of Data}, SIGMOD '16, pages 1119--1134, 2016.

\end{thebibliography}
\end{small}

\appendix
\section{Composing Relational and Graph Operators}
\label{sec:ComposingOperators}
In this section, we elaborate on how the design of \ourSys{} allows the composition of relational and graph operators in a query plan. The paper mainly focuses on performing graph operations on the leaf-level of a query execution pipeline and translating the results of the graph operations to a relational format. However, by introducing a new operator that creates temporary graphs from relational sub-queries, a query plan in \ourSys{} can have a graph operator as a non-leaf operator in a QEP. Thus, arbitrary composition of graph and relational operators in the same QEP becomes possible. To illustrate, we discuss all the possible compositions of the two types of operators, 
the relational operators 
and the graph operators. We ignore the relational-to-relational combination as it is a clear case supported natively in a typical RDBMS.

\subsection{Relational to Graph}
\label{R_To_G}
In all the previous examples, we 
have focused 
on query plans where the graph operators 
have been at
the leaf level of a QEP to process a 
user-defined 
graph view.
In many cases, it is important to define a graph view on the fly.
In essence, a user-defined graph view is a translation of the vertexes and the edges in relational tables to a graph structure. However, through the \textit{R-to-G} composition, we address the general case, where a graph operator can operate on a graph-view that is created on-the-fly from a child query-plan sub-tree, i.e., a temporary graph-view created by a SQL sub-query. To illustrate, consider Query $Q_{RG}$ in Listing~\ref{lst:R_To_G}. Query $Q_{RG}$ acts on a hospital database, where each patient has a location on a road network. The vertexes and the edges of the road network are stored in the \textit{Locations} and the \textit{Roads} relational tables, respectively. Query $Q_{RG}$ assumes further that each patient has a set of road types stored in the
\textit{UAvoidance}
table that denotes the road types 
to be
avoided by the patients (e.g., Patient~X avoids highways). The semantic of Query~$Q_{RG}$ in English reads as follows: "For each patient, say X, in San Francisco, find the email, the account id of Patient~X as well as the shortest path from Patient X to the hospital with a specific address, say 'Address~1', where the shortest path computation should not consider any road type 
to be
avoided by Patient X". For each patient selected by the outer query in Listing~\ref{lst:R_To_G}, there is a temporary graph-view created by a correlated sub-query. The \textit{TEMPGRAPH} keyword builds a temporary graph using a syntax similar to the \textit{Create Graph View} statement illustrated in Section~\ref{sec:CreateGraphView}. The temporary graphs are correlated with the patient tuples from the outer query to exclude the avoided road types. Notice that the paths of a temporary graph-view can be referenced normally in the \textit{Where-Clause} of the main query. For instance, the start vertex and the end vertexes in Query~$Q_{RG}$ are identified from the \textit{Patient}, and the \textit{Locations} tables, respectively. At query evaluation, the input to the \textit{TEMPGRAPH} operator is the relational inputs defining the vertexes and the edges of the temporary graph view, and the output is a materialized temporary graph-view that can be processed by a graph operator (which is the shortest path physical-operator in Query~$Q_{RG}$). Optimizing queries similar to Query~$Q_{RG}$ is an interesting future work.

\begin{lstlisting}[
      language=graphRelationalLang,
      showspaces=false,
      basicstyle=\ttfamily,
      numbers=none,
      caption={Constructing Temporary Graphs from Relational Sub-Queries},label={lst:R_To_G}]
SELECT Patient.Name, Patient.EMail, PS.PathString
FROM Patient, Locations Dest,
TEMPGRAPH(
VERTEXES (ID = LocId) FROM Locations
EDGES(ID = rId, FROM = rStart, TO = rEnd, Distance = rDist) FROM Roads WHERE Roads.Type NOT IN (SELECT Type FROM UAvoidance WHERE UAvoidance.UserId = Patient.Id)).Paths PS HINT(SHORTESTPATH(Distance))
WHERE PS.StartVertex.Id = Patient.LocationId AND PS.EndVertex.Id = Dest.Id
AND Dest.Address = "Address 1"
AND Patient.City = "San Francisco"
\end{lstlisting}

\subsection{Graph to Relational}
\label{G_To_R}
In this composition, the output of a graph operator can be digested by a relational operator.
As illustrated in Section~\ref{sec:HybridQEP}, \ourSys{} allows this composition by translating the output of a graph operator to a relational format (i.e., tuple data structure). Hence, a relational operator can process the graph operator's output as it flows in the QEP as relational tuples. 

\subsection{Graph to Graph}
\label{G_To_G}
In this composition, a graph operator can act on a sub-graph selected from an existing graph view, say $G_{base}$. This is possible in \ourSys{} through filtering predicates on the edges and the vertexes of the $G_{base}$ graph view, where the filtering predicates define the sub-graph to process by a graph operator. For instance, Listing~\ref{lst:G_To_G} computes the shortest path from Vertex~1 to Vertex~10, on the sub-graph formed by the roads with speed limit above 30~MPH in the RoadNetwork graph-view.

\begin{lstlisting}[
      language=graphRelationalLang,
      showspaces=false,
      basicstyle=\ttfamily,
      numbers=none,
      caption={A Graph Operator Acts on a Sub-Graph of a Graph View},label={lst:G_To_G}]
SELECT PS.PathString
FROM RoadNetwork.Paths PS HINT(SHORTESTPATH(Distance))
WHERE PS.Edges.SpeedLimit > 30 AND PS.StartVertex.Id = 1 AND PS.EndVertex.Id = 10
\end{lstlisting}

\section{Scaling on Cluster Environments}
\label{sec:ScalingOnCluster}
The \ourApproach{} approach can scale on a cluster environment by laying out the graph data on the cluster nodes. There is a wide spectrum of design choices for the graph layout on a cluster, e.g., the vertexes data and the edges data can be replicated or partitioned on the cluster nodes. Moreover, partitioning the vertexes or the edges relational-sources can be horizontal or vertical. Evaluating and optimizing the different approaches that emerge from the aforementioned choices is the focus of a future work.

One interesting approach of scaling on a cluster is to replicate only the graph topology (i.e., the graph view), and to allow arbitrary partitioning of both the vertexes and/or the edges relational-sources (e.g., horizontal partitioning based on a vertex's or an edge's attribute).
The main advantage of this approach is that graph traversals will be evaluated by traversing a local graph topology, however, other cluster nodes may be contacted to retrieve attributes of specific vertexes or edges (i.e., no hops to traverse other subgraphs on remote cluster nodes, but read requests are possible). Hence, doing expensive traversal requests on remote machines or moving the traversal state to a remote machine will be avoided.
Moreover, this approach is practical in many scenarios as the graph topology is typically much smaller than the vertexes and the edges attributes (i.e., graph data). Hence, replicating the graph topology may not be a significant overhead especially if the graph is used for analytics (i.e., read only workloads).
Notice that partitioning a graph is orthogonal to how the system processes distributed queries.
We realized this approach in \ourSys{} by implementing an interface that allows a cluster node to retrieve any attribute of a given vertex or edge during query evaluation, where each edge or vertex in the replicated graph-topology has a label that identifies the cluster machine hosting that edge or vertex. In essence, this approach is similar to Figure~\ref{Fig:graphViewMemory} except the existence of an additional indirection mechanism that allows a vertex or edge in a replicated graph-topology to retrieve any of its attributes from either the main memory if it is local, or from a remote machine if the attributes are hosted remotely.
Evaluating and optimizing this approach (e.g., doing eager bulk reads to reduce network communications), and comparing it to the aforementioned approaches is out of scope of this paper and will be the focus of a future work.

\section{Beyond Graph-Traversal Queries}
\label{sec:BeyondGraphTraversal}
Although graph traversal queries form a large body of vital graph queries, other queries like PageRank or user-defined graph algorithms may be important to many applications. It is clear that the \ourApproach{} approach can support new graph operations (e.g., PageRank) by adding new graph operators, however, it is more practical to allow executing user-defined graph operations without modifying the system internals.
This can be achieved by empowering the users with a mean to create procedural graph-operations, where the users can reference the graph-views singleton objects managed by the database through an API. This is possible in the \ourApproach approach as the graph views are native database objects that can be referenced for example by user-defined-functions (similar to manipulating a relational table in a UDF).

Introducing graph-aware user defined functions in the \ourApproach{} approach is beneficial as applications may have graph operations that require complex logic to be defined by the user. For example, the user may need to define a graph function to compute complex graph operations that are not provided by built-in functions (e.g., kNN, max-flow).
We highlight here a useful extension to \ourSys{}, where users can define two types of user-defined functions (UDFs) that are graph-aware, namely, scalar graph UDF, and graph-valued UDF.
Scalar graph UDFs are similar to relational scalar UDFs (e.g., Min, Max) and can appear in the Select-Clause except that graph-aware scalar UDFs can operate on graph views to compute a scalar value. For example, a user can create a graph-aware scalar UDF that computes the local clustering coefficient of a given vertex (e.g., \textit{Select LocalClusteringCoeff (Users.UserId)~\dots}).
In contrast to the scalar graph UDFs that return scalar values, the graph-valued UDFs return temporary graphs, and can be referenced in the From-Clause of a select statement, i.e., as if they are graph views. For example, Query~$Q_{RG}$ in Listing~\ref{lst:R_To_G} can be simplified by creating a graph-valued UDF, say \textit{ConstructTailoredRoadNetwork}, to construct a graph view that excludes specific road types from the base RoadNetwork based on the avoidance-preferences of a given patient identifier (i.e., \textit{ConstructTailoredRoadNetwork} can be parametrized by the patient identifier). The graph-aware UDFs can be written using an imperative language (e.g., C++), and registered in the database system so that users can reference them in declarative queries.

\section{Shortest-Path Queries with Filtering Predicates}
\label{sec:ExpShortestPaths}

In this section, we conduct an experiment using the Tiger road network to 
assess
the performance of \ourSys{} in evaluating the single-source shortest-path query (or SSSP, for short) in contrast to Grail~\cite{Grail_CIDR_2015}. The purpose of this experiment is to show that a simple algorithm, e.g.,
Dijkstra's algorithm~\cite{Dijkstra_1959} executing inside a relational database system can achieve significant performance gains over a pure-relational approach,
e.g., 
as in Grail~\cite{Grail_CIDR_2015}, when evaluating SSSP queries, or more generally, intensive traversal queries. Notice that the computational model of Grail is based on the vertex-centric computational approach 
that
is different from the graph-traversal model of \ourSys{}. However, both approaches have a common ground due to using an RDBMS in the evaluation. We
implement
the SSSP query of Grail as reported in Listing~3 in 
Grail's paper~\cite{Grail_CIDR_2015}. Our Grail implementation is an in-memory implementation on top of VoltDB to mitigate the disk IO cost, and we 
allow Grail to filter the edges while processing to report the effect of sub-graph selections on the query-execution performance.

\begin{figure}[h]
\centering
\includegraphics[width=2.24in]
{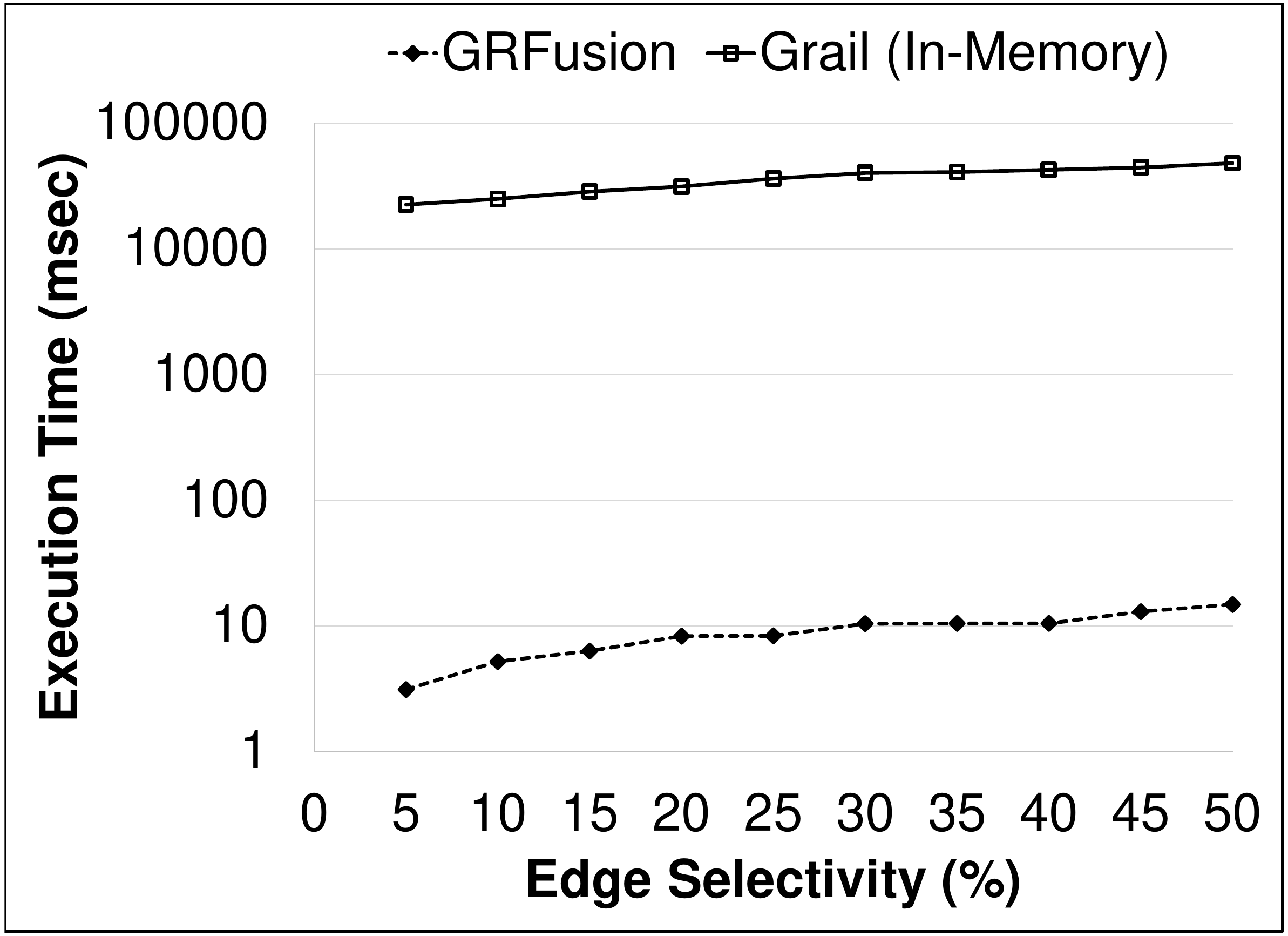}
\caption{\ourSys{} executes SSSP queries natively inside an RDBMS few-thousand times faster than Grail.}
\label{Fig:Q23_SSSP}
\end{figure}

We generate $1000$ random sources from which we execute an SSSP query to all the other vertexes, and we report the average query execution time for 
various 
sub-graph selectivity factors. Figure~\ref{Fig:Q23_SSSP} 
gives 
the performance of evaluating SSSP queries on the Tiger road network, where the x-axis and the y-axis are the edge-selectivity of the queries and the query-processing time in milliseconds, respectively. \ourSys{} achieves more than three orders-of-magnitude query-time speedup w.r.t. Grail. Notice that we do not use an advanced SSSP evaluation method. Instead, we use a straightforward Dijkstra's algorithm 
that utilizes
efficient filtering-predicates of the relational database engine. This emphasizes the point that having a native and an efficient graph representation inside an RDBMS can fill the gap between the RDBMSs and the graph algorithms that are designed for native graph structures, where these graph algorithms can achieve significant performance gains when 
compared
to equivalent pure-relational query evaluation approaches.

\end{document}